\documentclass{article}
\usepackage{blindtext}
\makeatletter
\renewcommand\paragraph{\@startsection{paragraph}{4}{\z@}%
            {-2.5ex\@plus -1ex \@minus -.25ex}%
            {1.25ex \@plus .25ex}%
            {\normalfont\normalsize\bfseries}}
\makeatother
\usepackage{titlesec}    
\titleformat{\chapter}[display]
{\Large%
    \Large
    \bfseries}{\chaptertitlename\ \thechapter}{10pt}{%
    \Large 
    }
\usepackage{amssymb}
\usepackage{lscape}
\usepackage{amsmath}
\usepackage{bm}
\usepackage[utf8]{inputenc}
\usepackage[T1]{fontenc}
\usepackage{mathtools}
\usepackage[thinc]{esdiff}
\usepackage{graphicx}
\usepackage{float}
\usepackage{array}
\usepackage{booktabs}
\usepackage[utf8]{inputenc}
\usepackage[utf8]{inputenc}
\usepackage{booktabs, makecell, longtable}
\usepackage{lipsum}
\usepackage{ltablex}

\usepackage{array}
\usepackage{pgfplots}
\restylefloat{figure}
\usepackage{mwe}
\usepackage{inputenc}
\usepackage[all]{xy}
\usepackage{breqn}

\usepackage{cancel}
\usepackage{color}
\usepackage{relsize}
\usepackage[margin=1in]{geometry} 
\usepackage{subcaption}
\usepackage{tikz}
\usetikzlibrary{calc,trees,positioning,arrows,fit,shapes,calc}
\usepackage{multirow,booktabs}
\usepackage{xcolor}
\usepackage{enumerate}
\newlength{\overwritelength}
\newlength{\minimumoverwritelength}
\newcommand{\overwrite}[3][red]{%
  \settowidth{\overwritelength}{$#2$}%
  \ifdim\overwritelength<\minimumoverwritelength%
    \setlength{\overwritelength}{\minimumoverwritelength}\fi%
  \stackrel
    {%
      \begin{minipage}{\overwritelength}%
        \color{#1}\centering\small #3\\%
        \rule{1pt}{9pt}%
      \end{minipage}}
    {\colorbox{#1!50}{\color{black}$\displaystyle#2$}}}

\newcommand*{\pd}[3][]{\ensuremath{\frac{\partial^{#1} #2}{\partial #3}}}

\newcommand\HUGE{\@setfontsize\Huge{38}{47}}
\begin{document}

\begin{center}
{\textbf{\Large{A Combined Experimental and Mathematical Study of The Evolution of Microbial Community Composed of Interacting \textsl{Staphylococcus} Strains}}}\\

$$\text{\bf{Nouf Alghamdi}}^1, \text{\bf{Mal Horsburgh}}^2, \text{\bf{Bakhtier Vasiev}}^1$$
$$\small{^{1}\,\text{Department of Mathematical Sciences, University of Liverpool, Liverpool, L69 7ZL, UK}}$$
$$\small{^{2}\,\text{Instititute of Integrative Biology, Biosciences Building, University of Liverpool, Crown Street,}}$$
$$\small{\text{Liverpool L69 7ZB, UK}}$$
\end{center}
\line(10,0){450}\\
{\textbf{\section*{Abstract}}}

The emergence of the phenomenon known as ABR (anti-bacterial resistance), is the result of the gradual decrease in the efficacy of antibiotics and the increase in the cost of producing new antibiotics. Hence, alternative solutions to prevent the spread of the pathogenic species are required.
%
%
Here we present a combined experimental and mathematical study of the evolution of microbial communities. The aim was to investigate the role of skin bacteria invasion and competition in limiting pathogenic species growth and colonisation, and to determine and reveal factors and conditions that alter and influence the dynamics of interactions between species. The focus in this study was \textsl{Staphylococcus aureus} as it is considered a major human pathogen that shows colonisation traits distinct from the more abundant skin antimicrobial-secreting residents, \textsl{S. epidermidis} and \textsl{S. hominis}. The method adopted when conducting this study was based on two approaches: experimental and mathematical.
The novelty and significance of this study lies in the fact that, unlike that found in a previous studies the manipulation of spatial structures, the level of toxicity, and initial frequencies did not prevent the emergence of resistance in the evolved \textsl{S. aureus} populations. The evolved \textsl{S. aureus} populations were able to dominate their opponents regardless of the environmental conditions. However, it was found that the level of toxicity and environmental regulations made it harder for evolved \textsl{S. aureus} populations to recover. 
\line(10,0){450}\\
{\textbf{\section{Introduction}}}

Over the years, the capacity to develop medicines to treat bacterial infections has resulted in significant progress in terms of reducing mortality rates \cite{ECDC}. The ability of bacteria to evolve, mutate, and reclaim control of the resident microbiome, on the other hand, demands the creation of alternative methods and approaches \cite{Gaur}.

The importance of the microbiota in avoiding the colonisation and proliferation of the pathogens is increasingly recognised \cite{Guarner F, Kamada N}. Most of the mechanisms for this beneficial effect of probiotic bacteria are indirect and include modification of the immune system, improvement of the intestinal epithelial barrier, or competition with pathogens for nutrients \cite{Bermudez Brito, Gourbeyre P, Kamada N, Macpherson A}. Bacteriocin proteins, which can kill pathogenic bacteria, are produced by several probiotic strains, and it has been demonstrated that an inhibitory generating \textsl{Escherichia coli} strain reduces colonisation by related pathogenic bacteria in the inflamed stomach of mice \cite{Sassone Corsi}. There is currently no proof that these systems are significant or common in humans. However, it is widely asserted that a probiotic diet improves human health \cite{Piewngam P, Sassone Corsi}.

Several studies indicate that \textsl{S. epidermidis} and \textsl{S. aureus}, which are the two dominant species in the nasal microbial community \cite{Wos Oxley ML, Yan M}, have negatively correlated distributions across nasal communities, suggesting that these species participate in one-way or mutual exclusion \cite{Frank DN, Libberton, P, Wos Oxley ML}. 

This method was demonstrated in the successful treatment of antibiotic-resistant bacteria that had been inhibited and restrained in growth when the host patient experienced a transplant from a healthy individual, helping to restore the beneficial resident bacteria by reproducing their abundance in the infected microbiome. When using the described population interactions, both bacterial strains will attempt to remain in a symbiotic environment, where they can transfer through all stages of their planned development until eventually collapsing with minimum disturbance. When a resident species of bacteria acts in its own habitat in isolation, referred to in biological terminology as intra-specific competition, it aims to maintain the same level of growth regardless of any interaction from other species. If preyed upon, it must also have the essential characteristics, whether it be the ability to generate or suppress toxins or the potential to mutate against them, to restore and sustain its size. Interactions and continuous competitions among different bacterial species involve many complex aspects.


Therefore, it was determined that to model the inhibitory interactions, it was necessary to introduce a new set of experiments to better investigate the hypothesis, which assumes that the interactions between bacterial communities limit the colonisation of pathogenic bacteria \cite{A}, and to conduct these interactions over a more extended period until these evolved populations converge toward a particular point or the change in their density is no longer significant. Thus, the study presented here aims to:
\begin{itemize}
\item Re-perform the experiments presented in \cite{A} that involve inhibitory interactions under mixed environmental conditions to understand the biological aspects and experimental techniques better and answer the questions raised in the previous study \cite{A}. 
\item Perform a series of experiments involving the selected species before engaging them in competitions to determine their characteristic features.
\item Examine inhibitory production and resistance evolution in invasion and competition under mixed conditions.
\item Extend the duration of the interactions to investigate the behaviour of the evolved species.
\item Perform a set of experiments involving the selected species after engaging them in the competitions to determine to what extent the interactions changed their characteristic features.
\item Develop mathematical models to explain and simulate the behaviour of evolved species when cultured separately, as well as explain and simulate the dynamics of interactions under mixed conditions.
\item Fit model parameters to experimental data. 
\item Perform numerical experiments to test and validate the three and four-variable model hypotheses.
\end{itemize}

{\textbf{\section{An Overview of Species Used and Experiments Performed in This Study}}}
Obtaining mathematical models capable of simulating the dynamics of these interactions, considering the differences in the conditions governing these interactions, must be achieved through a deep understanding of the biological aspects.
\begin{table}[ht]
\center
\begin{tabular}{|c|c|c|}
\hline
\begin{minipage}[t]{0.2\columnwidth}\raggedright
\center{Species} 
\end{minipage}
&
\begin{minipage}[t]{0.35\columnwidth}\raggedright
\center{Strain identifications}
\end{minipage}
&
\begin{minipage}[t]{0.35\columnwidth}\raggedright
\center{Inhibitor}
\end{minipage}\\
\hline
\begin{minipage}[t]{0.2\columnwidth}\raggedright
\center{\textsl{S. aureus}} 
\end{minipage}
&
\begin{minipage}[t]{0.35\columnwidth}\raggedright
\center{SH1000}
\end{minipage}
&
\begin{minipage}[t]{0.35\columnwidth}\raggedright
\center{Non-producing}
\end{minipage}\\
\hline
\begin{minipage}[t]{0.2\columnwidth}\raggedright
\center{\textsl{S. epidermidis}} 
\end{minipage}
&
\begin{minipage}[t]{0.35\columnwidth}\raggedright
\center{B180}
\end{minipage}
&
\begin{minipage}[t]{0.35\columnwidth}\raggedright
\center{Uncharacterised antibiotic}
\end{minipage}\\
\hline
\begin{minipage}[t]{0.2\columnwidth}\raggedright
\center{\textsl{S. epidermidis}}
\end{minipage}
&
\begin{minipage}[t]{0.35\columnwidth}\raggedright
\center{B155}
\end{minipage}
&
\begin{minipage}[t]{0.35\columnwidth}\raggedright
\center{Epifadin \cite{Ghabban}}
\end{minipage}\\
\hline
\begin{minipage}[t]{0.2\columnwidth}\raggedright
\center{\textsl{S. epidermidis}} 
\end{minipage}
&
\begin{minipage}[t]{0.35\columnwidth}\raggedright
\center{TU3298}
\end{minipage}
&
\begin{minipage}[t]{0.35\columnwidth}\raggedright
\center{The lantibiotic epidermin \cite{TU3298}}
\end{minipage}\\
\hline
\end{tabular}
\caption{\textbf{Identification of strains used in this study.}}
\label{tab:strains}
\end{table}
Close observation of the nature of these interactions would provide the opportunity to produce a model that could be considered a reflection of the dynamic processes involved therein and in making practical predictions.
\begin{table}[H]
\center
\begin{tabular}{|p{7.5cm} |p{7.5cm} | }
\hline
The experiment & Purpose of the experiment\\
\hline
Incubating replicates for each strain at $37^\circ C $\,for $24$ h, and taking $OD_{600}$ readings at $30$-min intervals.
& To determine the growth rate and generation time of bacteria.
\\
\hline
Incubating replicates for each strain in Petri dishes at $37^\circ C $\,for $24$ h, and taking the measurements of the spot in each strain before and after incubating. 
& To determine the rate of change in the spot size which is known as the diffusion coefficients.
\\
\hline
The growth inhibition assay, experimental study of toxin-mediated inhibition, pre-interactions.
& To determine the sensitivity of \textsl{\textsl{S. aureus}} strain against the toxins produced by \textsl{\textsl{S. epidermidis}} populations.
\\
\hline
Interactions, (competitions and inhibitions)
& To observe the dynamics of these interactions under different conditions.
\\
\hline
The growth inhibition assay, after interactions
& To evaluate the adaptations behaviour developed by \textsl{\textsl{S. aureus}} strain against the toxins produced by \textsl{\textsl{S. epidermidis}} populations.
\\
\hline
\end{tabular}
\caption{\textbf{A brief overview of experiments conducted and presented in this study.}}
\label{tab:experiments}
\end{table}
Here in this study, the possibilities of producing such a model by experimentally and theoretically defining such microbial interactions will be explored. In other words, both experimental and theoretical studies that cover different aspects of single and multi-species populations evolutions will be presented before presenting and introducing detailed explanations of these studies. (Table\,\ref{tab:strains}) presents an illustration of strains used in this study.

\begin{figure}[H]
\centering
\begin{subfigure}{0.48\linewidth}
\centering
\caption{}
\label{SH1000 growth curve}
\includegraphics[width=1\textwidth]{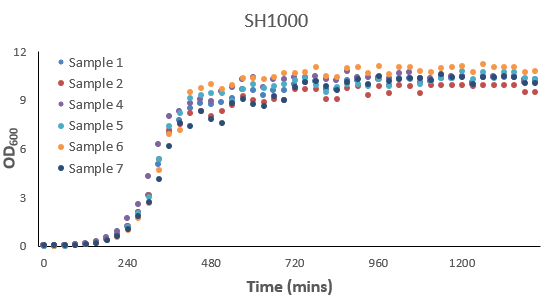}
\end{subfigure}
\hfill
\begin{subfigure}{0.48\linewidth}
\centering
\caption{}
\label{B180 growth curve}
\includegraphics[width=1\textwidth]{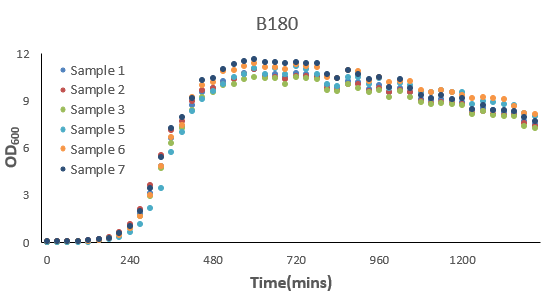}
\end{subfigure}
\\
\begin{subfigure}{0.48\linewidth}
\centering
\caption{}
\label{B155 growth curve}
\includegraphics[width=1\textwidth]{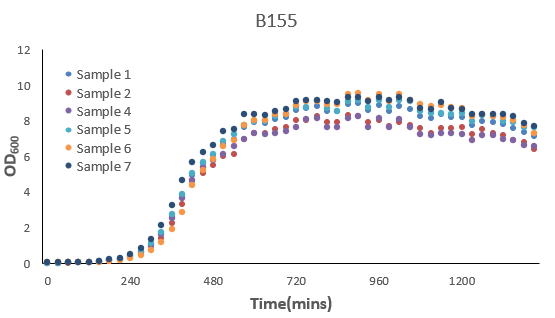}
\end{subfigure}
\hfill
\begin{subfigure}{0.48\linewidth}
\centering
\caption{}
\label{TU3298 growth curve}
\includegraphics[width=1\textwidth]{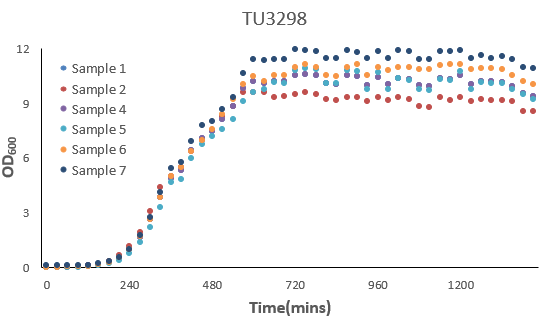}
\end{subfigure}
\hfill
\caption{\textbf{Growth curves.}\,Experimental data on the dynamics of growing populations. Panel (a): growth curves of \textsl{S. aureus} (SH1000) are represented by the dotted lines. Panel (b): growth curves of \textsl{S. epidermidis} (B180) are represented by the dotted lines. Panel (c): growth curves of \textsl{S. epidermidis} (B155) are represented by the dotted lines. Panel (d): growth curves of \textsl{S. epidermidis} (TU3298) are represented by the dotted lines. Dots represent the $OD_{600}$ readings, which were taken at $30$-min intervals. The $x$-axis is the time in minutes, and the $y$-axis is the optical density at $600\,nm$ $(OD_{600})$.}
\label{Eight replicates}
\end{figure}

Furthermore, (Table\,\ref{tab:experiments}), demonstrates a brief illustration and overview of experiments conducted and performed in Dr. Horsburgh’s laboratory in the Institute of Integrative Biology, University of Liverpool.
{\textbf{\section{Bacterial Growth Rate Dynamics}}}
Bacteria were frequently cultured in $10\,ml$ Brain Heart Infusion (BHI) broth (LabM) in a $20\,ml$ glass universal tube. Cultures were grown at $37^\circ C$, shaken at $200\,rpm$, revolutions per minute. Overnight cultures were typically grown for around $18 - 24\,h$. Exact details of the media can be found in (Table \ref{tab:media}). Strain stocks were preserved by adding either $700\,\mu l$ of an overnight culture, or a single colony collected from a plate and resuspended in $700\,\mu l$ of BHI broth (LabM) to $300\,\mu l$ of $50$ (v/v),  volume per volume glycerol and freezing at $-80^\circ C$. Duplicates were made of all freezer stocks.
\begin{table}[H]
\center
\begin{tabular}{ | p{7.5cm} | p{7.5cm} |}
\hline
Media / Buffers / Antibiotic & Composition \\
\hline
BHI agar plates & $3.7\%$ (w/v) BHI Broth (Lab M), $1.5\%$ (w/v), weight per volume
Agar-(Lab M), distilled water,\,$dd\,H_2O$\\
\hline
BHI broth & $3.7\%$ (w/v) BHI Broth (Lab M), $dd\,H_2O$ \\
\hline
Mannitol salt agar & $10.8\%$ (w/v) Mannitol Salt Agar (labM), $dd\,H_2O$\\
\hline
PBS & $0.8\%$ (w/v) $NaCl$, $0.034\%$ (w/v) $KH_2PO_4$, $0.12\%$ (w/v) $K_2HPO_4$\\
\hline
\end{tabular}
\caption{\textbf{Components for reagents used throughout this study.}}
\label{tab:media}
\end{table}
{\textbf{\subsection{Experimental Techniques}}}
Overnight cultures of each strain in Table (\ref{tab:strains}) were incubated for $24$ h in Growth Profiler 960 device, which generates growth curves of up to $960$ microbial cultures in microtiter plates. Because oxygen-transfer rates are readily reached (at $225 rpm/50mm$), exponential growth occurs up till $OD_{600}$ values of $3 -10$ (depending on the strain and its specific oxygen demand), which allows an accurate determination of maximal growth rates, see example curves at Fig (\ref{Eight replicates}). The shaker unit can be set to slow down (e.g., every $30$ minutes) to $30 rpm$ for a few seconds.  In these few seconds the culture comes temporarily to a rest (surface becomes close to horizontal), and the $10$ cameras make photos of the bottom of the (transparent) wells. Image analysis software in this device quantifies the cell density, and produces growth curves for all 960 strains, in this case, the biomass levels are expressed in $OD_{600}$ equivalents. 

Eight replicates for each strain were incubated after fixing the conditions that might influence their growth. For instance, the initial density, the abundance of the resources, temperature, speed of the shaker unit, reading interval and run time. $OD_{600}$ readings were taken at $30$-min intervals for all the involved spices.

\begin{figure}[H]
\centering
\begin{subfigure}{0.48\linewidth}
\centering
\caption{}
\label{}
\includegraphics[width=1\textwidth]{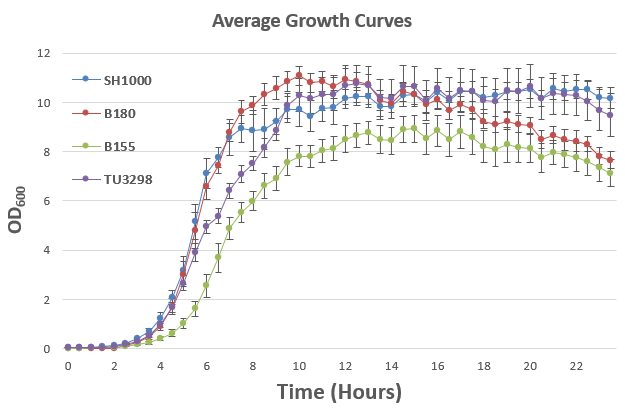}
\end{subfigure}
\hfill
\begin{subfigure}{0.48\linewidth}
\centering
\caption{}
\label{}
\includegraphics[width=1\textwidth]{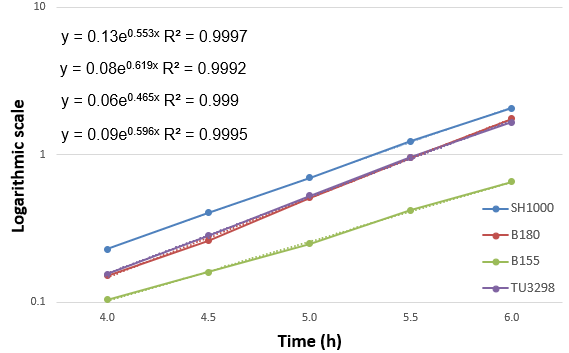}
\end{subfigure}
\caption{\textbf{Averaged growth curves of the involved strains.}\,Panel (a): illustrations of the bacterial growth curves after averaging the curves presented in Fig (\ref{Eight replicates}). The error bars show the standard deviation of the mean  $(n = 6)$. Panel (b): depicts the exponential phase used to calculate the doubling time of cultured populations. The $y$ equations represent the line equations by changing the $y$-axis into logarithmic scale, $R$ denotes the R-squared values. The $x$-axis is the time in hours, and the $y$-axis is the optical density at $600\,nm$ $(OD_{600})$.}
\label{growth curves 2}
\end{figure}

{\textbf{\subsection{Experimental Results}}}
As illustrated in Fig (\ref{growth curves 2}),\, the growth of bacterial communities entails four primary stages: lag phase, log phase, stable or stationary phase and death phase where the size of the population starts to decline.
After liquid culture broths are inoculated for the involved species, the proliferation of bacteria does not start immediately but takes some time to proliferate. As stated earlier, the time between inoculation and the beginning of multiplication is known as the lag phase. In this phase, the inoculated bacteria become familiar with the environment, activate various enzymes, and adapt to the environmental temperature and surrounding conditions. During this phase, there is an increase in the size of bacteria but no visible increase in the number of bacterial cells. The cells function metabolically. Since all the surrounding conditions for all the incubated species are fixed, the lag phase duration varies according to the bacterial species.

As shown in Fig (\ref{Eight replicates}), it can be determined that all inoculated species consumed approximately the same amount of time in this phase, which is four hours, taking into account that \textsl{S. aureus} population, SH1000, Fig (\ref{SH1000 growth curve}), were the fastest and \textsl{S. epidermidis} population, B$155$, Fig (\ref{B155 growth curve}) were the slowest. 

Subsequently, the start of a new phase known as the log phase was noted, characterised by rapid exponential cell growth, where the bacterial population doubles during every generation. They increased at their maximum rate. The growth rates of B180, Fig (\ref{B180 growth curve}), and TU3298, Fig (\ref{TU3298 growth curve}), populations were the greatest during this phase, while B155 population scored the least growth rates at this phase. Since the rapidly dividing cultures were not provided with constant addition of nutrients and frequent removal of waste products, this phase was brief for all incubated species. As shown in Fig (\ref{growth curves 2}), the log phase appears as a steeply sloped straight line.

After the log phase, the bacterial growth almost stopped entirely, due to lack of necessary nutrients, lack of water and oxygen, changes in pH of the medium, and accumulation of their own toxic metabolic wastes. This phase is known as the stationary phase. It was during this phase that the cultures were at their greatest population density. However, the death rate of bacteria exceeded the rate of reproduction of bacteria, as is rapidly evident particularly in B180 population. The last phase in the growth curve is known as the decline phase. During this phase, the bacterial population declined due to the death of cells. The death rate of B180 populations was the greatest during this period, while SH1000 populations maintained their stability for a longer time.  

{\textbf{\subsection{Mathematical Analysis of The Obtained Results}}}
From the graphs obtained in the laboratory of $OD_{600}$ against time over $24$ hours, the sampling time can be identified that occurred during the exponential phase of the growth curves and by using exponential curve fitting function, equations were generated. The equations take the following form:
\begin{equation}
\displaystyle y=\displaystyle y_0\,e^{\displaystyle rt},
\label{R}
\end{equation}
\noindent where $y$ is the concentration of the population, $t$ is the time, $y_{0}$ is the value of $y$ at time $0$ and $r$ is the the growth rate.
By re-arranging this equation, it was possible to estimate the doubling time, relaxation time, and growth rate for all the involved species per minute, hour, and day as follows:
$$
\text{Relaxation time}=\,\displaystyle\frac{t_2-t_1}{B},
$$
$$
\text{Growth rate}=\,\displaystyle\frac{1}{\text{Relaxation time}}=\frac{B}{t_2-t_1},
$$
$$
\text{Doubling time}=\,\displaystyle\frac{\ln2}{\text{Growth rate}},
$$
where $t_1$ and $t_2$ are two consecutive time points throughout the bacterial growth, and $B$ is a positive constant representing the relative growth rate.

Conducting these experiments enabled the obtaining of accurate data and readings regarding the growth rate and doubling time for the involved populations by using the logarithmic scale for the log phase for all evolved populations, as seen in Fig (\ref{growth curves 2}b). The averaged curve for each strain was obtained from the best six replicates results, Fig (\ref{growth curves 2}a). According to (Table \ref{tab:doubling time}), there is no significant difference in terms of the doubling times nor the growth rates between any of the \textsl{S. epidermidis} strains tested and the \textsl{S. aureus} strain SH1000 used in this study.

%

\begin{table}[H]
\center
\resizebox{\textwidth}{!}{\begin{tabular}{|c|c|c|c|c|c|c|c|c|c|c|c|}
\hline
\multicolumn{1}{|c|}{\multirow{2}{*}{Species}}
&\multicolumn{1}{c|}{\multirow{2}{*}{Relative rate $B$}}
&\multicolumn{3}{c|}{Doubling time}&
\multicolumn{3}{c|}{Relaxation time}&
\multicolumn{3}{c|}{Growth rate}\\
\multicolumn{1}{|c|}{}&\multicolumn{1}{c|}{}&\multicolumn{1}{c}{Min}&\multicolumn{1}{c}{Hour}&\multicolumn{1}{c|}{Day}&\multicolumn{1}{c}{Min}&\multicolumn{1}{c}{Hour}&\multicolumn{1}{c|}{Day}&\multicolumn{1}{c}{Min}&\multicolumn{1}{c}{Hour}&\multicolumn{1}{c|}{Day} \\
\hline
SH1000
& $0.553$ 
&$37.623$
&$0.627$
&$0.026$
&$54.279$
&$0.905$
&$0.038$
&$0.018$
&$1.105$
&$26.530$\\
\hline
B180
& $0.619$ 
&$33.583$
&$0.560$
&$0.023$
&$48.450$
&$0.807$
&$0.034$
&$0.021$
&$1.238$
&$29.722$\\
\hline
B155
& $0.465$ 
&$44.690$
&$0.745$
&$0.031$
&$64.475$
&$1.075$
&$0.045$
&$0.016$
&$0.931$
&$22.334$\\
\hline
TU3298
& $0.596$ 
&$34.884$
&$0.581$
&$0.024$
&$50.327$
&$0.839$
&$0.035$
&$0.020$
&$1.192$
&$28.613$\\
\hline
\end{tabular}}
\caption{\textbf{Doubling times and growth rates of strains used in this study.}\,The doubling times were compared to SH1000 (\textsl{S. aureus}) as a control using a post hoc Dunnett’s test \cite{Everett B}.}
\label{tab:doubling time}
\end{table}

{\textbf{\subsection{Mathematical Models and Simulations}}}
Population evolution growth rate can be independent of the size of the population. This only has a chance of being successful while the resources available to the population are unlimited. If these resources are finite and limited, as in the situation here, then there is a maximum population size that can be supported by the environment. This maximum population is often called the carrying capacity and it is proportional to the abundance of resources, $K\,\propto\,R$. In these circumstances, the growth rate must depend on the size of the population and specifically it must approach zero as the population approaches the carrying capacity. In this way it is possible to arrive at the logistic model for population growth: 
$$
\displaystyle\frac{dN}{dt}= r\,N\left(1\,-\,\frac{N}{R}\right),
$$
where $r$ and $R$ are positive constants representing the linear growth rate and the abundance of resources, respectively. For small $N$ (compared with $R$) the growth rate $r(N)$ is close to the linear growth rate $r$. This is consistent with the fact that when the population is small, resources are plentiful and seemingly infinite. It is only when the population size enlarges that the effect of the finiteness of resources is noticeable. This model was first examined by the Belgian Mathematician, Pierre Francois Verhulst, in the middle of the $19^{th}$\, century, \cite{K}. In this model, as in the exponential model, the only mechanisms for changing the population size are births and deaths; there is nothing to account for migration into or out of the population. However, by adding another variable that represents the resources as following:
\begin{equation}
\displaystyle\frac{dR}{dt}= (a- s\,R) -\,cN.
\label{resource}
\end{equation}
In the absence of $N$ , the resource $R$ is depleted due to natural factors at a rate $s\,R$, while it is replenished to a stable level $a$. At equilibrium, the rate of production equals the rate of depletion. Hence, the first term is equal to zero. When $N$ is present, the resource is additionally depleted at a rate $cN$ , where $c$ is known to be the consumption rate of the existing population. The equation (\ref{resource}) represents the general case when a single population consumes a single resource \cite{Tilman}. However, as these experiments were incubated for only $24$ hours, and during this period no food was introduced or added for the incubated populations besides the initial amount that was presented at the beginning of this experiment, this means that $a = 0$ and the reduction of the resources due to natural factors will not be significant when compared to the decline caused by population consumption. 

Hence, only the population consumption will noticeably influence the evolution of the resource equation. This means the values of the first term in the resource equation, production, and depletion due to other reasons, will not be significant compared to the second term, which represents the reduction of limited resources due to population consumption. Thus, the first term of this equation can be neglected, and the system of equations will take the form: 
\begin{equation}
\begin{cases} 
\displaystyle\frac{dN}{dt}= r\,N\left(1\,-\,\frac{N}{R}\right),\\[8pt]
\displaystyle\frac{dR}{dt}= -\,cN,
\end{cases}
\label{EQ:growth model}
\end{equation}
where $N = N(t)$ represents the densities of the populations and $R = R(t)$ represents the resource abundances at time $t$ and $c$ represents positive constant rates of consumption. Initial conditions $N(0)$ for all species were obtained and defined in the laboratory by defining the $OD_{600}$, which is an abbreviation indicating the optical density of a sample measured at a wavelength of $600\,nm$, before incubating the samples. Furthermore, growth rates $r$ were determined for each incubated population from Table (\ref{tab:doubling time}).
\begin{figure}[H]
\centering
\begin{subfigure}{0.24\linewidth}
\centering
\caption{}
\label{Simulation of growth curve SH1000}
\includegraphics[width=1\textwidth]{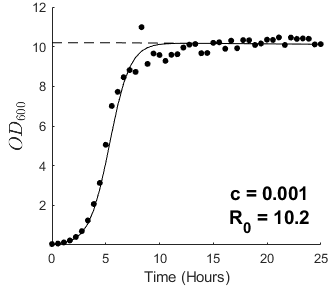}
\end{subfigure}
\hfill
\begin{subfigure}{0.24\linewidth}
\centering
\caption{}
\label{Simulation of growth curve B180}
\includegraphics[width=1\textwidth]{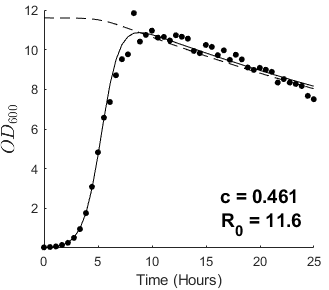}
\end{subfigure}
\hfill
\begin{subfigure}{0.24\linewidth}
\centering
\caption{}
\label{Simulation of growth curve B155}
\includegraphics[width=1\textwidth]{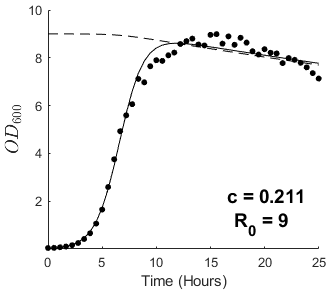}
\end{subfigure}
\hfill
\begin{subfigure}{0.24\linewidth}
\centering
\caption{}
\label{Simulation of growth curve TU3298}
\includegraphics[width=1\textwidth]{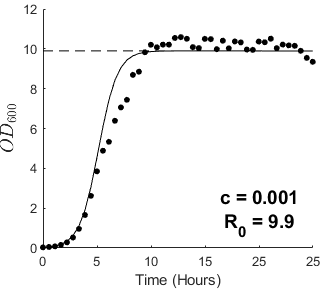}
\end{subfigure}
\hfill
\caption{\textbf{Modelling the growth of bacterial populations.}\,The simulations of growth curves (solid lines) and resource evolution (dashed lines) generated from \eqref{EQ:growth model} plotted against the actual experimental data (black dots). Panel (a): represents SH1000. Panel (b): B180. Panel (c): B155. Panel (d): TU3298. Different values of $c$’s indicate different rates of resource consumption, whereas different values of $R_0$ indicate different initial values of the limited resource. The $x$-axis is the time in hours, and the $y$-axis is the optical density at $600\,nm$ $(OD_{600})$.}
\label{Simulation of growth curve}
\end{figure}
According to the model (\ref{EQ:growth model}) simulations, and as shown in Fig (\ref{Simulation of growth curve B180}), the highest rate of resource consumption was achieved by \textsl{S. epidermidis} population, B180, with $c = 0.461$ and initial resource density $R_0 = 11.6$, followed by \textsl{S. epidermidis} population, B155, Fig (\ref{Simulation of growth curve B155}), with $c = 0.211$ and initial resource density $R_0 = 9$. According to figures (\ref{Simulation of growth curve TU3298}) and (\ref{Simulation of growth curve SH1000}), both \textsl{S. epidermidis}, TU3298, and \textsl{S. aureus}, SH1000 populations ranked third with the same consumption rates $c = 0.001$ and initial resource density $R_0 = 9.9$,\, $R_0 = 10.2$, respectively. 

{\textbf{\subsection{Determining The Diffusion Coefficients of The Involved Species}}}
To determine the diffusion coefficients, all strains involved were cultured on BHI agar plates before experiments. Bacteria were cultured for 24 h on $10\,cm$\,diameter BHI agar plates, and the lawns of \textsl{S. aureus} (SH1000) and \textsl{S. epidermidis} strains Table (\ref{tab:strains}) were then scraped off the agar plates and suspended in $10 \,mL$ of PBS by vortexing thoroughly. The $cfu/mL$ in each tube was equalized by diluting the cell suspensions in PBS and comparing the $OD_{600}$ of each suspension (approximately $5 \times 10^8\,cfu/mL$ for \textsl{S. aureus} and \textsl{S. epidermidis}, determined by viable count). All isolates were vortexed thoroughly before $50 \mu L$ (containing approximately $2.5 \times 10^6$ cells) was plated onto $25 mL$ BHI agar and incubated at $37^\circ C$. Measurements of the cultured spots were taking for each plate prior to incubation and after, i.e., measurements of initial and incubated spots.
\begin{figure}[H]
\centering
\includegraphics[width=0.7\textwidth]{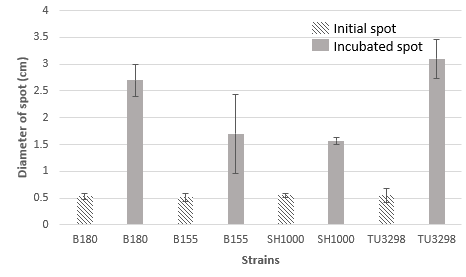}
\caption{\textbf{Diagram showing the size of the spot diameter before and after the incubation.}\, Dashed bars indicate the average diameter of the bacterial spot before incubation; solid bars are the same indications after the incubation pointed out for B180, B155, SH1000, and TU3298, respectively in the $x$-axis. $y$-axis is the diameter of the spot in (cm). Error bars represent the standard error of the mean.}
\label{spot diameter}
\end{figure}
According to the diagram shown in Fig\,(\ref{spot diameter}), the average size of the spot diameter for all the involved strains before incubating is approximately $0.5\, cm$, where the size of the diameter for the incubated spots varies from one strain to another. It is worth noting that \textsl{S. epidermidis} (TU3298) recorded the highest rate of diffusion while \textsl{S. aureus} (SH1000) ranked last in the rate of diffusion. Experimentally, it was possible to define the rate of change in the size of the incubated spots by finding the ratio between the average size of the diameter of incubated spots and initial spots, as follows: 
\begin{equation}
\displaystyle\frac{B180(24)}{B180(0)}=5.06,\,\,\displaystyle\frac{B155(24)}{B155(0)}=3.29,\\
\displaystyle\frac{SH1000(24)}{SH1000(0)}=2.85,\,\,\displaystyle\frac{TU3298(24)}{TU3298(0)}=5.64.
\label{ratio of spot}
\end{equation}
Mathematically, it was possible to estimate the diffusion coefficient for each strain by rescaling and adapting the actual sizes of medium and spots into the simulation. As noted previously, the size of Petri dishes used in the experiments is $10\,cm$ and the average size of the diameter of pre-incubated spots for all strains is $0.5\,cm$. Therefore, $L = 1$ was chosen as the size of the medium; $hx = L/n$ as the space step size, where $n = 100$ the number of grid points. Furthermore, a column vector was constructed for each strain that represented the initial conditions. The initial vectors were divided into $100$ grid points and the spot was introduced to these zero vectors as non-zero values in the middle where $a = 0.05$ indicates the diameter of the initial spots for all strains i.e., 5 space steps. Thus, the ratios between the length of the medium and the diameter of the initial spot were maintained the same in the simulations as in the actual experiments: 
$$
\displaystyle\frac{\text{Diameter of the initial spot}}{\text{Diameter of the medium}} =\displaystyle \frac{0.5}{10} = \displaystyle\frac{0.05}{1} = 0.05.
$$
Since the diffusion coefficient for each strain was estimated by culturing them independently for a day, this can be represented mathematically by the logistic equation \eqref{EQ:363}.
\begin{equation}
\displaystyle u_t= D\,u_{xx}+ r\,u\left(1-u\right),
\label{EQ:363}
\end{equation}
where $r$ and $D$ are positive parameters representing the growth rate and the diffusion coefficient respectively. 
In order to obtain the diffusion coefficients for the involved strains, it is necessary to define the proper values for $D$, which is the diffusion coefficients in simulations, that satisfy the actual experimental ratio between the average size of diameter of incubated spots and initial spots presented in (\ref{ratio of spot}). 

The diffusion coefficient for each strain was obtained by considering the ratio between the initial spot and 24-hour incubated spot diameters obtained in \eqref{ratio of spot} as conditions and by starting to introduce different values for the diffusion coefficient, using loops in MATLAB, it was possible to determine the diffusion coefficients for the involved strains that satisfy the obtained ratios in \eqref{ratio of spot} with minimum error.
Thus: 
\begin{equation}
D_{B180}=2\times10^{-5}, D_{B155}=5\times10^{-6}, D_{SH1000}=1.8\times10^{-6}, D_{TU3298}=5\times10^{-5}.
\label{diffusion coefficients1}
\end{equation}
where $D_{B180} = 2\times10^{-5}$ in our simulation space and time units corresponds to $D_{B180} = 2\times10^{-3}\,\displaystyle\frac{cm^2}{day}$ in the actual experimental units, and similarly for the diffusion coefficients of other strains.
\begin{figure}[H]
\centering
\includegraphics[width=1.\textwidth]{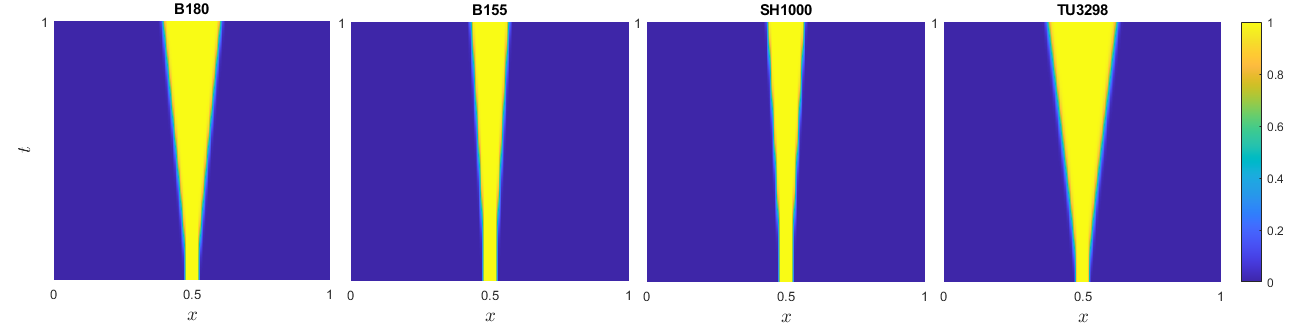}
\caption{\textbf{Mathematical simulation of the incubated spot evolution.} Surface plots showing the one-day evolution of the incubated spots for the involved species, B180, B155, SH1000, and TU3298, respectively. The $x$-axis represents space ($10\,cm$), the $y$-axis represents time (1 day), the colour bar indicates the cell density.}
\label{diffusion coefficients}
\end{figure}
As demonstrated in Fig (\ref{diffusion coefficients}), which shows simulations obtained from (\ref{EQ:363}) representing the evolution of the incubated spots for the populations (B180), (B155), (SH1000), and (TU3298), respectively.

In terms of the toxins, according to \cite{1}, the characteristic diffusion constant for a molecule the size of a monomeric protein is $\approx\,100 \mu m^2/s$ in water and is about ten-fold smaller, $\approx\,10 \mu m^2/s$, inside a cell. Hence rescaling this coefficient to the units used in these simulations yields:
$$
D_{\text{toxin}}=\displaystyle\frac{100\times10^{-8}}{1.16\times10^{-5}}\,= 8.6\,\times 10^{-2}\,\displaystyle\frac{cm^2}{day}\, \Leftrightarrow 8.6\,\times 10^{-4}\,\displaystyle\frac{s.s.u}{s.t.u}.
$$
where $s.s.u$ is the simulation space unit and $s.t.u$ means the simulation time unit.

{\textbf{\section{Study of Toxin-mediated Inhibition}}}
As noted previously, the aims of these experiments were to determine the sensitivity of \textsl{S. aureus} strain against the toxins produced by \textsl{S. epidermidis} populations. Hence, this type of experiment was performed twice. Firstly, the parental isolate of all \textsl{S. epidermidis} species was used to determine and measure the effect of their toxins on the involved \textsl{S. aureus} populations before starting the invasions. Secondly, the parental isolate was used at the end of performed interactions, to investigate the evolution of resistance by the evolved populations of \textsl{S. aureus}. This was achieved when ancestral and evolved \textsl{S. aureus} strains were sprayed over ancestral and evolved \textsl{S. epidermidis}, toxin-producing residents. 

{\textbf{\subsection{Experimental Techniques}}}

As indicated in (Table \ref{tab:strains}), three independent \textsl{S. epidermidis} isolates were selected, two of them from the previous study that sampled the anterior nares of $60$ healthy volunteers (Libberton et al. 2014). The following method was adapted to test competitor strains with inhibitor- producing strains. 

\begin{figure}[H]
\centering
\begin{subfigure}{0.49\linewidth}
\centering
\caption{}
\label{inhibition assay - before 1}
\includegraphics[width=1\textwidth]{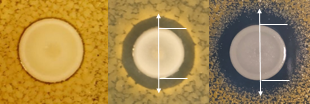}
\end{subfigure}
\\
\begin{subfigure}{0.49\linewidth}
\centering
\caption{}
\label{inhibition assay - before 2}
\includegraphics[width=1\textwidth]{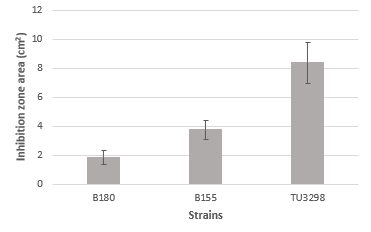}
\end{subfigure}
\hfill
\begin{subfigure}{0.49\linewidth}
\centering
\caption{}
\label{comparison of spots diameter1}
\includegraphics[width=1\textwidth]{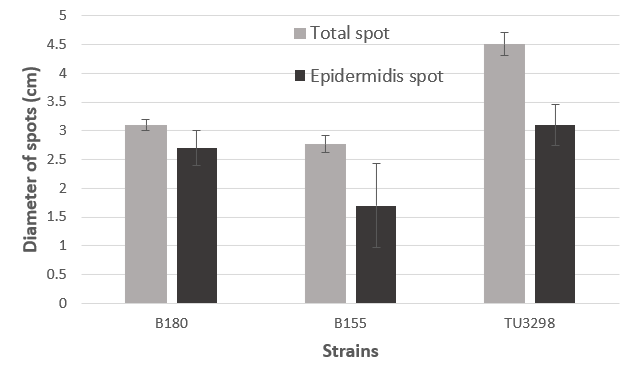}
\end{subfigure}
\caption{\textbf{Results of toxin-mediated inhibition, (pre-invasions).}\,Panel (a): examples of the actual inhibition zone produced by the ancestral \textsl{S. epidermidis} B180, B155, and TU3298 strains against the ancestral SH1000, respectively. Panel (b): the inhibition zone area ($cm^2$) produced by the ancestral inhibitor-producing \textsl{S. epidermidis} strains, pointed in the $x$-axis, against the ancestral SH1000. Panel (c): the diameters of both \textsl{S. epidermidis} spots and total spots, i.e., the inhibition circle + \textsl{S. epidermidis} spot, for all three strains. Error bars represent the standard error of the mean.}
\label{comparison of spots diameter}
\end{figure}

The inhibition spray assay was based on the protocol described previously in \cite{Inhibition}, by using SH1000 as the indicator strain. A $25\,\mu l$\,\,spot (approximately $10^8$ cells) of an overnight bacterial culture was pipetted onto the centre of an agar plate containing $15\,ml$ of BHI agar (lab M). The plates were incubated for $18\,h$ at $37^\circ C$\,before $250\,\mu l$ of a ten-fold diluted overnight culture of \textsl{S. aureus} SH1000 ($10^6\,cfu$) was sprayed over the plate. The plates were incubated again for a further $18\,h$, when the size of the inhibition zone produced by the central nasal isolate on SH1000 was assessed. The experiment was repeated 5 to 10 times to obtain accurate and consistent results.
The clarity of the inhibition zone was scored according to a simple scoring system of 1 to 4, 4 being completely clear and 1 being no detectable zone. The areas of any detectable zones were also recorded by measuring the diameter of the inhibition zone and the central colony. The area of both the zone and the colony were calculated using the equation: 
$$ \displaystyle Area= \pi r^2,$$
where $r$ is the radius of the colony or the inhibition zone. The central colony area was then subtracted from the total zone area, leaving only the area of the zone around the perimeter of the central colony.
{\textbf{\subsection{Experimental Results (Pre-invasions)}}}
As shown in Fig (\ref{inhibition assay - before 1}), all three isolates were toxin producers as revealed in a deferred inhibition assay by their killing of \textsl{S. aureus} [zone of clearing] when a lawn of \textsl{S. aureus} strain, SH1000, was sprayed over them. 

Although one of the species, B180, produces an inhibition zone against \textsl{S. aureus}, in the study presented in this thesis, these particular isolates were considered non-toxin producers based on their not significantly reducing the viability of strain SH1000 when comparing the obtained zone of inhibition with the other isolates, B155 and TU3298, [see Fig (\ref{inhibition assay - before 2})]. 

Moreover, SH1000 displayed no growth inhibition activity against any of the selected \textsl{S. epidermidis} strains when performing the mutual deferred inhibition assay. Of the two toxin-producing \textsl{S. epidermidis} strains, TU3298 produced an inhibition area that was around two times greater than that of B155.

\begin{figure}[H]
\centering
\begin{subfigure}{0.49\linewidth}
\centering
\caption{}
\label{inhibition assay - after 1}
\includegraphics[width=1\textwidth]{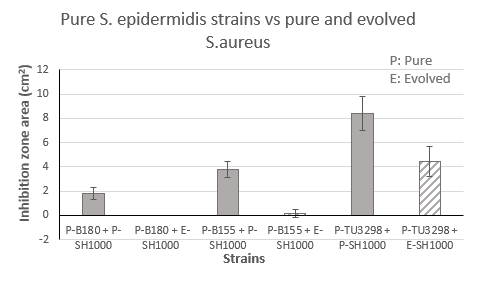}
\end{subfigure}
\hfill
\begin{subfigure}{0.49\linewidth}
\centering
\caption{}
\label{inhibition assay - after 2}
\includegraphics[width=1\textwidth]{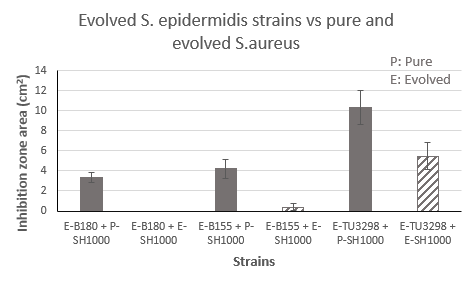}
\end{subfigure}
\hfill
\begin{subfigure}{0.49\linewidth}
\centering
\caption{}
\label{comparison of spots diameter2}
\includegraphics[width=1\textwidth]{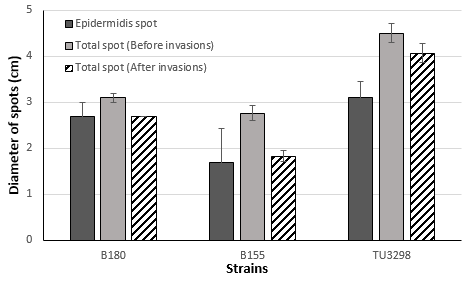}
\end{subfigure}
\caption{\textbf{Resistance of SH1000 before and after competing against \textsl{S. epidermidis} (B180, B155 and TU3298).}\,Panel (a): shows inhibition zones produced by pure \textsl{S. epidermidis} strains, and panel (b): by the evolved \textsl{S. epidermidis} strains against the pure SH1000 (P) and the evolved SH1000 (E). In both panels, the $y$-axis represents the inhibition zone area ($cm^2$). Panel (c): for each strain, a comparison of the diameter of the inhibition zone before (light solid bars) and after invasions (dashed bars), as well as the diameter of the \textsl{S. epidermidis} spot (dark solid bars). Error bars represent the standard error of the mean.}
\label{Inhibition Assay1}
\end{figure}

{\textbf{\subsection{Experimental Results (After-invasions)}}}
According to the performed competition outcomes, \textsl{Staphylococcus aureus} populations were able to restrict, inhibit, and invade communities of \textsl{S. epidermidis} under mixed conditions. To test whether the evolution of inhibitory toxin resistance by \textsl{S. aureus} was responsible for the inhibition and invasion in a mixed environment, evolved \textsl{S. aureus} strains were sprayed over ancestral and evolved \textsl{S. epidermidis}, toxin-producing residents. In most cases, evolved \textsl{S. aureus} were resistant to the \textsl{S. epidermidis} toxins (see Fig \ref{Inhibition Assay1}). 

Furthermore, the size of the inhibition zones produced by ancestral and evolved TU 3298 against evolved \textsl{S. aureus} decreased, indicating that it is only a matter of time before the evolved \textsl{S. aureus} populations completely resist the toxins produced by this strain, as there is a positive association between the level of toxicity expressed by \textsl{S. epidermidis} and the time consumed by \textsl{S. aureus} populations to adapt to this toxicity.

%
%

Interestingly, as seen in Fig (\ref{Inhibition Assay1}, solid bars), evolved \textsl{S. epidermidis} populations had greater inhibitory activity on ancestral SH1000, Fig(\ref{inhibition assay - after 2}), than ancestral \textsl{S. epidermidis} populations, Fig(\ref{inhibition assay - after 1}). 

This supports the opinion that the toxin producer developed to overcome the survival challenges forced by increasingly resistant \textsl{S. aureus} populations. The evolved \textsl{S. epidermidis} may have increased the production of the inhibitory toxin or started the production of different toxins. 

{\textbf{\subsection{Mathematical Models and Simulations}}}
Two bacterial populations in a one-dimensional domain are considered. One species, the ‘producer,’ produces an antimicrobial toxin that inhibits the other, the susceptible. Denoting the concentrations of susceptible, producer and toxin at position $x$ and time $t$ by $u(x, t)$, $v(x, t)$ and $T (x, t)$, this ecosystem is modelled via the equations:
\begin{equation}
\begin{cases} 
\displaystyle\frac{\partial u}{\partial t} = D_u\frac{\partial^2 u}{\partial x^2}+r_u\,u\,(1-u-pT),\\[8pt]
\displaystyle\frac{\partial v}{\partial t} = D_v\frac{\partial^2 v}{\partial x^2}+r_v\,v\,(1-v),\\[8pt]
\displaystyle\frac{\partial T}{\partial t} = D_T\frac{\partial^2 T}{\partial x^2}+f_1\,v-f_2\,T,\\
\end{cases} 
\label{toxin model}
\end{equation}
where $u=u(x,t) , v=v (x,t)$ and $T=T(x,t)$ are concentrations of \textsl{S. aureus}, \textsl{S. epidermidis} and toxic substance. $p$: inhibition coefficient, $f_1$: inhibitor production rate, $f_2$: inhibitor degradation rate.\\
Initial conditions:
$$
\displaystyle u(x,0)= 0.1\,\,\,,\,T(x,0)=0 \,\,\forall\,x\,\in\,[0,L],\,\,\text{and}\,\,
v(x,0)=\begin{cases} 
\displaystyle 1\,\,\,\,\,\,\,\,\,\,\,\,\,\,\,x\in[\frac{L}{2}-l,\frac{L}{2}+l],\\
\displaystyle 0\,\,\,\,\,\,\,\,\,\,\,\,\,\,\,\text{otherwise}.\\
\end{cases}
$$
Boundary condition;\\
Again, zero–flux boundary conditions are imposed for all variables. 
\begin{equation}
u_x(0, t)= u_x(L, t)= v_x(0, t)= v_x(L, t)= T_x(0, t)= T_x(L, t)=0 \,\,\,\,\,\forall \, t >0.
\label{B.C1}
\end{equation}
Since the solution $T(x, t)$ represents a travelling wave, the structure of the solution will be the same for all time and the speed of spread of this shape is a constant, denoted by $c$. If this wave is considered in a travelling form moving at speed $c$, it will appear stationary. Mathematically we can say that if the solution 
\begin{equation}
T(x,t) = T(x-ct) = T(\zeta), \,\,\,\zeta = x-ct
\label{x-ct}
\end{equation}
then $T(x,t)$ is a travelling wave, and it moves at constant speed $c$ in the positive $x$-direction.
Rewriting the equation that represents the toxin substance in the system (\ref{toxin model}) by using the form of the solution presented in (\ref{x-ct}), gives:
\begin{equation}
\frac{\partial T}{\partial t} = D_T\frac{\partial^2 T}{\partial x^2}+f_1\,v-f_2\,T,
\label{toxins in chapter 3}
\end{equation}
$$
\frac{\partial T}{\partial t} = -cT\,', \,\, \frac{\partial T}{\partial x} = DT\,', \,\, \frac{\partial^2 T}{\partial x^2}= DT\,''.
$$
Thus, the equation (\ref{toxins in chapter 3}), takes this form:
\begin{equation}
DT\,''+cT\,'-f_2\,T=-f_1\,v.
\label{toxins in chapter 3_1}
\end{equation}
To be able to solve this equation certain assumptions should be made, such as that at equilibrium there is no movement, hence $c = 0$. Also, the value of $v$ which represents the concentration of the producer is equal to one inside the cultured spot, whereas it equals zero everywhere else, thus the equation (\ref{toxins in chapter 3_1}), takes the following form:
\begin{equation}
DT\,''-f_2\,T=\begin{cases} 
-f_1\,\,\,\,\,\,\,\,\,\,\,\,\,\,\,x\leq|a|,\\
0\,\,\,\,\,\,\,\,\,\,\,\,\,\,\,\text{otherwise}.\\
\end{cases}
\label{T equation}
\end{equation}
To find the general solution to the non-homogeneous differential equation (\ref{T equation}), in this case the solution can be written in this form:
\begin{equation}
T(x,t)= T_{cf}+T_{p},
\label{General solution}
\end{equation}
where $T_{cf}$ is the solution to the complementary function $DT\,''-f_2\,T=0$, and $T_p$ is the particular solution to the equation $DT\,''-f_2\,T=G(x)$. First we start by finding the solution for the complementary function, by setting:
$$
T=e^{mx}, \,\,\,\,\,\,\, T\,'=me^{mx}, \,\,\,\, T\,''=m^2e^{mx}.
$$
Inserting these values into the complementary function, yields:
\begin{equation}
Dm^2e^{mx}-f_2e^{mx}=0 \to Dm^2-f_2=0 \to m=\pm\sqrt{\frac{f_2}{D}}.
\label{value of m}
\end{equation}
Thus:
$$
T_{cf}= c_1 e^{mx} +c_2 e^{-mx}.
$$
Now, it is necessary to find the specific solution to the equation (\ref{T equation}), since the form of $G(x)$ is constant, meaning that $T_p=c$ and $T\,''=0$. Inserting these values into the equation (\ref{T equation}), yields:
$$
-f_2 c= -f_1 \to c=\frac{f_1}{f_2}.
$$
Thus, the equation (\ref{General solution}) becomes:
$$
T(x,t)= c_1 e^{mx} +c_2 e^{-mx} + \frac{f_1}{f_2}.
$$
Therefore, 
\begin{equation}
T(x,t)= \begin{cases} 
\displaystyle c_1 e^{mx} +c_2 e^{-mx} + \frac{f_1}{f_2}\,\,\,\,\,\,\,\,\,\,\,\,\,\,\,x\leq|a|,\\[8pt]
\displaystyle c_3 e^{mx} +c_4 e^{-mx}\,\,\,\,\,\,\,\,\,\,\,\,\,\,\,\,\,\,\text{otherwise}.\\
\end{cases}
\label{T solution}
\end{equation}
The values of the constants can be determined from the boundary conditions (\ref{B.C1}):\\
$T_x(0, t)=0$:\\
$$
\displaystyle c_1 me^{m(0)} -c_2 me^{-m(0)}=0 \,\,\to c_1 m-c_2 m=0 \,\,\,\to c_1=c_2.
$$
$T_x(L, t)=0$:\\
\begin{equation}
\displaystyle c_3 me^{m(L)} -c_4 me^{-m(L)}=0 \,\,\to c_3 e^{m L}=c_4 me^{-m L} \,\,\,\to c_3=c_4 e^{-2m L}.
\label{c3}
\end{equation}
In order to define these constants, we need to add two more conditions as follows:
$\lim_{x \to a^+} T(x, t)= \lim_{x \to a^-} T(x, t)$ and $\lim_{x \to a^+} T_x(x, t)= \lim_{x \to a^-} T_x(x, t)$. From these two conditions it is possible to obtain values of the constants $c_1, c_3$ and $c_4$.
$$
\displaystyle c_{1}\,{\mathrm{e}}^{a\,m}+c_{1}\,{\mathrm{e}}^{-a\,m}+\frac{f_{1}}{f_{2}}=c_{4}\,{\mathrm{e}}^{-2\,L\,m}\,{\mathrm{e}}^{a\,m}+c_{4}\,{\mathrm{e}}^{-a\,m}.
$$
Solving this equation for the constant $c_1$, yields:
\begin{equation}
\displaystyle c_{1}= \frac{c_{4}\,{\mathrm{e}}^{-a\,m}+c_{4}\,{\mathrm{e}}^{-2\,L\,m}\,{\mathrm{e}}^{a\,m}-\frac{f_{1}}{f_{2}}}{{\mathrm{e}}^{a\,m}+{\mathrm{e}}^{-a\,m}}.
\label{C1-1}
\end{equation}
Applying the second condition, gives:
$$
\displaystyle c_{1}\,m\,{\mathrm{e}}^{a\,m}-c_{1}\,m\,{\mathrm{e}}^{-a\,m}=c_{4}\,m\,{\mathrm{e}}^{-2\,L\,m}\,{\mathrm{e}}^{a\,m}-c_{4}\,m\,{\mathrm{e}}^{-a\,m}.
$$
Again, solving this equation for the constant $c_1$, yields:
\begin{equation}
\displaystyle c_{1}=\frac{c_{4}\,{\mathrm{e}}^{2\,L\,m}-c_{4}\,{\mathrm{e}}^{2\,a\,m}}{{\mathrm{e}}^{2\,L\,m}-{\mathrm{e}}^{2\,L\,m}\,{\mathrm{e}}^{2\,a\,m}}.
\label{C1-2}
\end{equation}
From equations (\ref{C1-1}) and (\ref{C1-2}), we have:
\begin{equation}
\displaystyle \frac{c_{4}\,{\mathrm{e}}^{-a\,m}+c_{4}\,{\mathrm{e}}^{-2\,L\,m}\,{\mathrm{e}}^{a\,m}-\frac{f_{1}}{f_{2}}}{{\mathrm{e}}^{a\,m}+{\mathrm{e}}^{-a\,m}}=\frac{c_{4}\,{\mathrm{e}}^{2\,L\,m}-c_{4}\,{\mathrm{e}}^{2\,a\,m}}{{\mathrm{e}}^{2\,L\,m}-{\mathrm{e}}^{2\,L\,m}\,{\mathrm{e}}^{2\,a\,m}}.
\label{theC4}
\end{equation}
Solving the equation (\ref{theC4}) for $c_4$\,, gives:
$$
\displaystyle c_4=-\frac{f_{1}\,{\mathrm{e}}^{2\,L\,m}\,\left({\mathrm{e}}^{2\,a\,m}-1\right)}{2\,f_{2}\,\left({\mathrm{e}}^{a\,m}-{\mathrm{e}}^{m\,\left(2\,L+a\right)}\right)}.
$$
Inserting the value of $c_4$ into the equation (\ref{c3}), yields:
$$
\displaystyle c_3=-\frac{f_{1}\,\left({\mathrm{e}}^{2\,a\,m}-1\right)}{2\,f_{2}\,\left({\mathrm{e}}^{a\,m}-{\mathrm{e}}^{m\,\left(2\,L+a\right)}\right)}.
$$
Similarly for the equation (\ref{C1-2}):  
$$
\displaystyle c_1=\frac{f_{1}\,\left({\mathrm{e}}^{2\,L\,m}-{\mathrm{e}}^{2\,a\,m}\right)}{2\,f_{2}\,\left({\mathrm{e}}^{a\,m}-{\mathrm{e}}^{m\,\left(2\,L+a\right)}\right)}.
$$
Thus, the solution (\ref{T solution}) takes the following form:
\begin{equation}
T(x,t)= \begin{cases} 
\displaystyle \frac{f_{1}\,\left({\mathrm{e}}^{2\,L\,m}-{\mathrm{e}}^{2\,a\,m}\right)}{2\,f_{2}\,\left({\mathrm{e}}^{a\,m}-{\mathrm{e}}^{m\,\left(2\,L+a\right)}\right)}\,(e^{mx} +e^{-mx}) + \frac{f_1}{f_2}\,\,\,\,\,\,\,\,\,\,\,\,\,\,\,\,\,\,\,\,\,\,\,\,\,\,x\leq|a|,\\[8pt]
\displaystyle -\frac{f_{1}\,\left({\mathrm{e}}^{2\,a\,m}-1\right)}{2\,f_{2}\,\left({\mathrm{e}}^{a\,m}-{\mathrm{e}}^{m\,\left(2\,L+a\right)}\right)} \,e^{mx} -\frac{f_{1}\,{\mathrm{e}}^{2\,L\,m}\,\left({\mathrm{e}}^{2\,a\,m}-1\right)}{2\,f_{2}\,\left({\mathrm{e}}^{a\,m}-{\mathrm{e}}^{m\,\left(2\,L+a\right)}\right)}\, e^{-mx}\,\,\,\,\,\,\,\,\,\,\,\,\,\,\,\,\,\,\text{otherwise}.\\
\end{cases}
\label{T_solution}
\end{equation}

\begin{figure}[H]
\centering
\begin{subfigure}{0.49\linewidth}
\centering
\caption{}
\label{T_profile}
\includegraphics[width=1\textwidth]{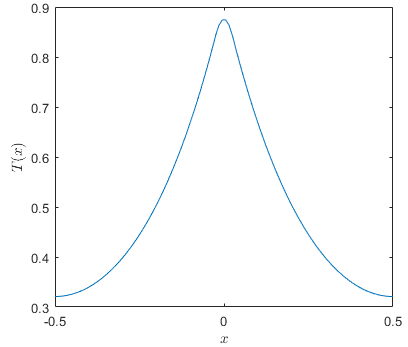}
\end{subfigure}
\hfill
\begin{subfigure}{0.49\linewidth}
\centering
\caption{}
\label{f_1 and f_2 curves}
\includegraphics[width=1\textwidth]{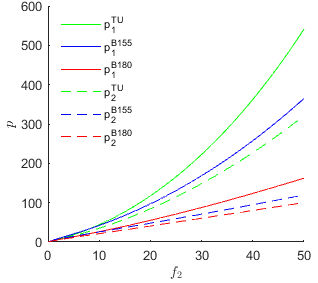}
\end{subfigure}
\hfill
\caption{\textbf{Profile of toxin and illustration of the relation between the inhibition coefficient $p$ and decaying rate $f_2$, Eq \eqref{f_1 and f_2 equation}.}\,Panel (a): illustration of the solution $T(x,t)$ in the equation (\ref{T_solution}). The $x$-axis represents space, and the $y$-axis represents toxin concentration. Parameters: $L=0.5,\,a=0.025,\,D=0.086,\,f_1=1,\,f_2=0.1$. Panel (b): solid lines represent the association of the inhibitory coefficients between the toxins and the susceptible fraction of \textsl{S. aureus}, whereas the dashed lines represent the same associations with the adapted fraction of \textsl{S. aureus} populations. TU3298 in green, B155 in blue, and B180 in red. Parameters: in solid lines, (green) $a=0.155$ and $x=0.225$, (blue) $a=0.085$ and $x=0.138$, and (red) $a=0.135$ and $x=0.155$. In dashed lines, (green) $x=0.203$, (red) $x=0.135$, and (blue) $x=0.092$.}
\end{figure}

After obtaining a graph of the solution $T(x,t)$ shown in Fig(\ref{T_profile}), the aim is to define a relationship between the parameters $p$, $f_1$ and $f_2$ which represent the inhibition coefficient, the production and the decaying of the toxin substance respectively. Since these parameters cannot be determined precisely, as in the case in determining the growth rate as well as the diffusion coefficients for the involved bacterial strains, the aim is to define a relationship between these coefficients based on what there is in the model (\ref{toxin model}).
In order to achieve this, the first equation in the system (\ref{toxin model}) is considered, at equilibrium $\dot u=0$, when $u=0 \,\,\text{or}\,\, u=1-pT$, from (\ref{T_solution}) this can be written as following:
$$
\displaystyle 1-p\,\Big(-\frac{f_{1}\,\left({\mathrm{e}}^{2\,a\,m}-1\right)}{2\,f_{2}\,\left({\mathrm{e}}^{a\,m}-{\mathrm{e}}^{m\,\left(2\,L+a\right)}\right)} \,e^{mx} -\frac{f_{1}\,{\mathrm{e}}^{2\,L\,m}\,\left({\mathrm{e}}^{2\,a\,m}-1\right)}{2\,f_{2}\,\left({\mathrm{e}}^{a\,m}-{\mathrm{e}}^{m\,\left(2\,L+a\right)}\right)}\, e^{-mx}\Big)=u.
$$
According to (\ref{T_solution}), this solution of $T$ covers the area outside of \textsl{S. epidermidis} spots, where the inhibition zone occurs and beyond, which means that $u$ has either vanished or is dying in this area. Therefore, for simplicity, we set $u=0$, and this yields:

%

$$
\displaystyle -\frac{f_1\,p}{f_2}\,\Big(\frac{{\mathrm{e}}^{2\,a\,m}-1}{2\,\,\left({\mathrm{e}}^{a\,m}-{\mathrm{e}}^{m\,\left(2\,L+a\right)}\right)} \,e^{mx}+\frac{{\mathrm{e}}^{2\,L\,m}\,\left({\mathrm{e}}^{2\,a\,m}-1\right)}{2\,\left({\mathrm{e}}^{a\,m}-{\mathrm{e}}^{m\,\left(2\,L+a\right)}\right)}\, e^{-mx}\Big)=1,
$$
\begin{equation}
\displaystyle f_1\,p=\frac{-f_2}{\Big(\frac{{\mathrm{e}}^{2\,a\,m}-1}{2\,\,\left({\mathrm{e}}^{a\,m}-{\mathrm{e}}^{m\,\left(2\,L+a\right)}\right)} \,e^{mx}+\frac{{\mathrm{e}}^{2\,L\,m}\,\left({\mathrm{e}}^{2\,a\,m}-1\right)}{2\,\left({\mathrm{e}}^{a\,m}-{\mathrm{e}}^{m\,\left(2\,L+a\right)}\right)}\, e^{-mx}\Big)},
\label{f_1 and f_2 equation}
\end{equation}
where $a$ represents the radius of the \textsl{S. epidermidis} spot, and $x$ is the radius of the total spot including the \textsl{S. epidermidis} spot plus the inhibition zone. As $f_1\,p$ is a constant, hence, for simplicity $f_1=1$ can be set to reduce the number of parameters when plotting the relationship between $p$ and $f_2$.

According to Fig (\ref{comparison of spots diameter1}), there are three different strains with three different values of $a$ and $x$. To obtain a better understanding of the relationship between $p$ and $f_2$, it was decided to plot $p$ against different values of $f_2$, and not the other way around as the value of $m$ in the equation (\ref{value of m}) depends on the value of $f_2$.

As seen in Fig (\ref{Inhibition Assay1}), the evolved population SH1000 mutated against the toxins produced by B180, as no inhibition zones were observed when applying evolved SH1000 on pure B180 nor evolved. As a result, when producing the red dashed line in Fig (\ref{f_1 and f_2 curves}), which shows the relationship between the inhibition coefficient and the decaying rate in an evolved mixed population of \textsl{S. aureus} and \textsl{S. epidermidis} B180, $a=x$ was set in the equation (\ref{f_1 and f_2 equation}), meaning that the radius of the total spot is equal to the radius of the B180 spot, zero inhibition zone, (See Fig \ref{comparison of spots diameter2}). Since it was not possible to quantify the inhibition coefficient and the decaying rates of the toxins produced by the involved \textsl{S. epidermidis} strains, i.e., the values of $p$ and $f_2$, the equation (\ref{f_1 and f_2 equation}) indicates the relationship between these factors. Therefore, whenever any of these values can be defined the other will lie within the lines plotted and shown in Fig (\ref{f_1 and f_2 curves}).
{\textbf{\section{Dynamics of Interacting Population}}}
Before exploring and showing the materials and the methods of the competitions, several concepts need to be explained. These concepts are presented as follows:\\
{\textbf{Types of Interactions}}

\begin{enumerate}[1.]
\item Competition for resources.\\
Competition for resources occurs when two or more species in a community compete for a shared resource. When one species consumes a finite resource, it leaves less available for other species and those that rely on it to suffer as a result. Species acting in their own best interests will utilise a finite resource until it is depleted, resulting in a population crash \cite{2}. Competitors might avoid this disaster by using resources more slowly \cite{4} or by using a different resource altogether, a phenomenon known as niche partitioning \cite{6}. \textsl{S. aureus} and \textsl{Pseudomonas aeruginosa} compete for iron scavenging resources in vitro \cite{3} and in vivo (rat infection model) \cite{5}. However, in the nasal environment, this has yet to be proven.
\item Inhibition and competition for resources.\\
When one organism produces a chemical that reduces the relative fitness of another organism in the community, it is referred to as toxin-mediated interference competition \cite{11}. Toxin-mediated interference is popular in bacteria from the manufacture of bacteriocins \cite{9}, antibiotics \cite{13}, and secondary metabolites \cite{10}. However, toxin-mediated interference has not been studied in connection to the nasal microbial population or colonisation. Although \textsl{Staphylococci} have a wide spectrum of bacteriocins capable of killing closely- related species \cite{8}. \textsl{S. epidermidis} produces two bacteriocins that have been rigorously studied and have been shown to kill \textsl{S. aureus}. These are \textsl{epidermin} \cite{7} and \textsl{gallidermin} in aureus \cite{12}.
\end{enumerate}
{\textbf{Environmental Structure}

Interference competition in bacterial communities is often induced by costly environmentally produced toxins and is thus likely to be influenced by spatial population structure \cite{15}. Based on this knowledge, the experiments were performed in two different environmental structures to investigate the influence of the environmental structure on the dynamics of interactions. These two structures were known as mixed and structured environments, and they were distinct in terms of transferring the populations, as shown in Fig (\ref{environmental structure}). In a mixed environment, the competing populations were introduced into a new and fresh medium every day by scraping the entire bacterial lawn off before thoroughly vortexing and then pipetting onto a new plate. In contrast, in a structured environment, the transfers were made by replica, plating with velvet to maintain spatial structure.
\begin{figure}[H]
\centering
\includegraphics[width=0.9\textwidth]{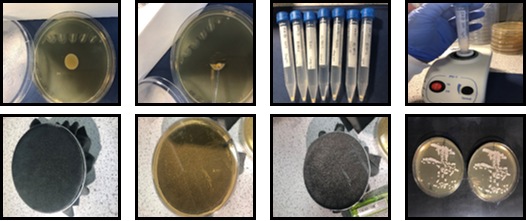}\\
\caption{\textbf{Illustrations of different types of environmental structure.}\,Panels in the first row show the steps of the daily transfer process among the populations interacting under mixed conditions. In contrast, the second row shows the actions taken under structured conditions.}
\label{environmental structure}
\end{figure}

{\textbf{Initial Frequencies}}

As illustrated previously, the initial frequency, along with the environmental structure, plays a significant role in determining the outcomes of these interactions and influencing the dynamics of interactions. Hence, different initial frequencies were considered to test this theory in the experiments in this study. Therefore, three different sets of experiments were performed. The first set concerned populations of toxin-producing \textsl{S. epidermidis} invading resident populations of susceptible \textsl{S. aureus} at a concentration of (0.01: 1). The second set involved what is defined as the mutual invasions, where populations of toxin-producing \textsl{S. epidermidis} were invaded by a susceptible population of \textsl{S. aureus} at a concentration of (1: 0.01). The third set involved performing these interactions from equal initial frequency (1:1).\\
{\textbf{Level of Toxicity}}

As shown in (Table\,\ref{tab:strains}), three different strains from the \textsl{S. epidermidis} family were used, distinguished by their level of toxicity, where (B180) is considered a low toxin strain, (B155) with a moderate level of toxicity and (TU3298) is a high toxin strain. These toxicity levels were determined as revealed in a deferred inhibition assay by their killing of \textsl{S. aureus} [zone of clearing when a lawn of \textsl{S. aureus} strain SH1000 was sprayed over them], (See Fig \ref{comparison of spots diameter}). Of the three toxin-producing \textsl{S. epidermidis} strains, B180 produced an inhibition area that was around three times smaller than that of B155. On the other hand, TU3298 produced an inhibition area that was around two times greater than that of B155.

Before the start of the invasion experiment, all nasal isolates were cultured on BHI agar plates. Bacteria were cultured for $18\,h$ on $100\,mm$ diameter BHI agar plates when the lawns of \textsl{S. aureus} (SH1000) and \textsl{S. epidermidis} strains (resident and invader – Table\,\ref{tab:strains}) were scraped from the agar plates and suspended in $10\,ml$ of PBS (Table\,\ref{tab:media}), (containing approximately $5 \times 10^8 cfu/ml$ for \textsl{S. aureus} and \textsl{S. epidermidis}, determined with a colony count) by vortexing thoroughly.
By diluting the cell suspensions in PBS and measuring the $OD_{600}$ of each suspension, the $cfu/ml$ in each tube was equalised. In a final volume of $10\,ml$ PBS, the two species were mixed with the invader at a different frequency (ratio) to the resident ($0.01:1$). The mixtures were well-vortexed before plating $50\,\mu l$ (containing about $2.5 \times 10^6$ cells) on $25\,ml$ BHI agar and incubating at $37^\circ C$. Five replicate communities were established at each starting frequency.

The communities were transferred to a new agar plate every day, the mixed environmental conditions explained in the previous section and shown in Fig (\ref{environmental structure}). Furthermore, each isolate determined viable counts by scraping the bacterial lawn from the plate and transferring it to $10 ml$ of sterile PBS. After thoroughly vortexing and then pipetting $50 \mu l$ onto a new plate to complete the experiment, the competing populations were counted by using the serial dilution method ($100 \mu l$ of the sample+$900 \mu l$ PBS) (See Fig \ref{Viable counts}). 

The colony-forming unit (CFU) is a microbiological counting unit used to quantify the number of viable microorganisms in a sample. The term ‘viable’ refers to microorganisms that can divide and are alive. Unlike other methods that count the number of cells regardless of viability, this approach counts just the living cells because this term reflects the number of bacteria capable of reproducing when colonies develop on the plate. A CFU calculation requires sampling \cite{16}. The viable counts were accomplished on the structured plates by scraping the remaining bacterial lawn after duplicate plating and serial diluting in PBS. Colony counts were done on BHI plates, and colony morphology and colour were used to distinguish colonies.
\begin{figure}[H]
\centering
\includegraphics[width=0.5\textwidth]{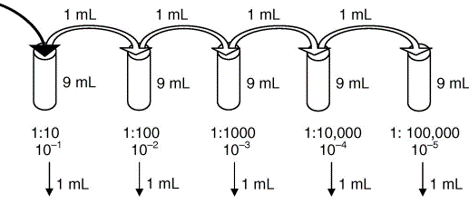}\\
\caption{\textbf{Serial dilution method.}\,A serial dilution is used to dilute a microbial sample enough to obtain single colonies when plating. $100 \mu l$ of liquid containing the bacteria is mixed with $900 \mu l$ of PBS liquid in order to give the dilution a 1:10 ratio, and several dilutions are frequently recommended. Taken from \cite{16}.}
\label{Viable counts}
\end{figure}
After obtaining the experimental data, i.e., the number of colonies in the fraction of the population, the CFU calculation in the original sample can be achieved using the equation \cite{16}:
\begin{equation}
\text{Number of CFU per $ml$}=\frac{\text{number of colonies}\times\text{dilution factor}}{\text{size of the sample($ml$)}}
\label{viable count equation}
\end{equation}
As explained earlier, there occurred two types of interactions. The first type involved resource competitions, interactions between \textsl{S. epidermidis} B180 and \textsl{S. aureus} SH1000, while the second type involved resource competitions and the produced toxins played a significant role in the outcomes of these invasions, interactions between \textsl{S. epidermidis} B155, TU3298, and \textsl{S. aureus} SH1000.

As shown in Fig (\ref{Fig:f10d}), the performed interactions lasted for varying periods until the change in the competing population density was no longer significant. Thus, the interactions between \textsl{S. aureus} strain, SH1000, and the inhibitor-producing \textsl{S. epidermidis} species, B155 and TU3298, were carried on for 28 days, which is longer than the interactions between SH1000 and the low-inhibitor-producing strain, B180 (17 days). 
{\textbf{\subsection{Experimental Results}}}
When performing the interactions under mixed conditions, the following findings were obtained:\\
{\textbf{Regardless of the initial concentrations and the level of toxicity, \textsl{S. aureus} populations were always able to limit the presence of their opponents.}}

Under mixed conditions, the pathogenic species SH1000 dominated all interactions performed regardless of the manipulated factors imposed (level of toxicity and initial frequency). Such findings are consistent with those of the study presented in \cite{A}, as it was demonstrated that \textsl{S. epidermidis} was never able to successfully invade under mixed conditions, and that \textsl{Staphylococcus aureus} was only able to invade toxin-producing \textsl{S. epidermidis} under mixed conditions.\\
 
\begin{figure}[H]
\centering
\begin{subfigure}{0.32\linewidth}
\centering
\caption{}
\label{C1}
\includegraphics[width=1.\textwidth]{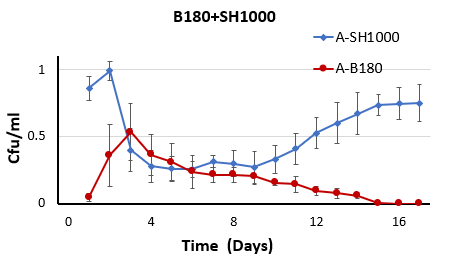}
\end{subfigure}
\hfill
\begin{subfigure}{0.32\linewidth}
\centering
\caption{}
\label{C2}
\includegraphics[width=1.\textwidth]{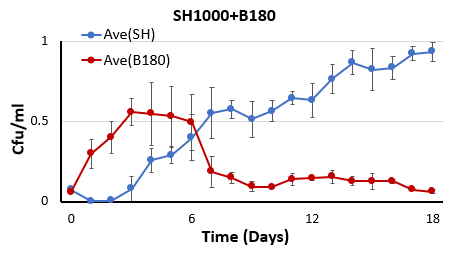}
\end{subfigure}
\hfill
\begin{subfigure}{0.32\linewidth}
\centering
\caption{}
\label{C3}
\includegraphics[width=1.\textwidth]{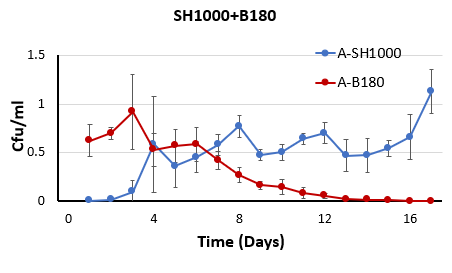}
\end{subfigure}
\\
\begin{subfigure}{0.32\linewidth}
\centering
\caption{}
\label{C4}
\includegraphics[width=1\textwidth]{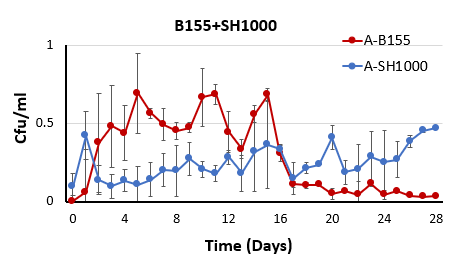}
\end{subfigure}
\hfill
\begin{subfigure}{0.32\linewidth}
\centering
\caption{}
\label{C5}
\includegraphics[width=1\textwidth]{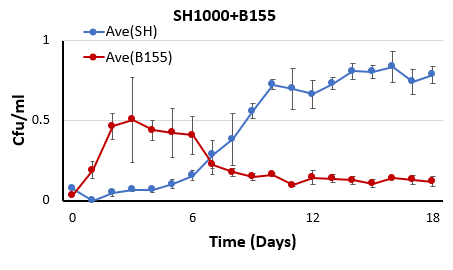}
\end{subfigure}
\hfill
\begin{subfigure}{0.32\linewidth}
\centering
\caption{}
\label{C6}
\includegraphics[width=1\textwidth]{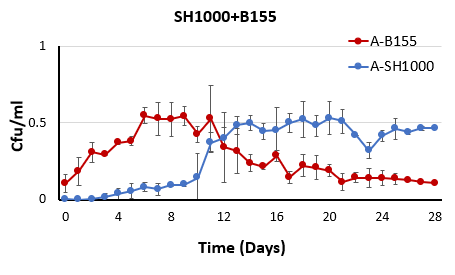}
\end{subfigure}
\\
\begin{subfigure}{0.32\linewidth}
\centering
\caption{}
\label{C7}
\includegraphics[width=1\textwidth]{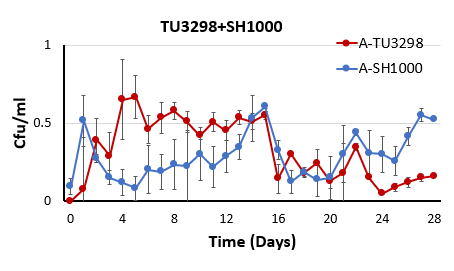}
\end{subfigure}
\hfill
\begin{subfigure}{0.32\linewidth}
\centering
\caption{}
\label{C8}
\includegraphics[width=1\textwidth]{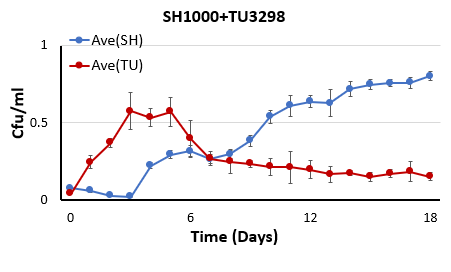}
\end{subfigure}
\hfill
\begin{subfigure}{0.32\linewidth}
\centering
\caption{}
\label{C9}
\includegraphics[width=1\textwidth]{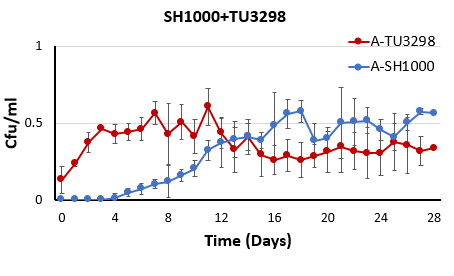}
\end{subfigure}
\caption{\textbf{The experimental data on the interaction dynamics of \textsl{S. epidermidis} and \textsl{S. aureus} species in mixed environments.}\, Panels (a), (b) and (c): show the interactions between low-toxin producing isolates of \textsl{S. epidermidis} (B180, red) and populations of \textsl{S. aureus} (SH1000, blue), starting from different initial concentrations. Panels (d), (e), and (f): reflect the interactions between moderate toxicity populations of \textsl{S. epidermidis}, (B155, red) and \textsl{S. aureus}, (SH1000, blue). Panels (g), (h), and (i): show the interactions between highly toxic populations of \textsl{S. epidermidis} (TU3298, red) and \textsl{S. aureus} (SH1000, blue). Panels in the first column, (a), (d), and (g): represent invasions of \textsl{S. epidermidis} (invaders) at initial ratios of (0.01: 1) to (resident) \textsl{S. aureus} populations. Panels in the second column, (b), (e) and (h): when the evolutions between the interacted populations started from equal initial frequencies. Panels in the third column, (c), (f) and (i): represent the mutual invasions performed between invaders of \textsl{S. aureus} at initial ratios of (0.01: 1) to resident populations of \textsl{S. epidermidis}. The $x$-axis is the time in days, and the $y$-axis is the colony-forming units (CFU) per plate. Error bars represent the standard error of the mean (n = 3).}
\label{Fig:f10d}
\end{figure}
{\textbf{A positive association between the interaction level of toxicity and the time consumed by the \textsl{S. aureus} population to recover.}}

During the interactions, all developed isolates of \textsl{S. aureus} displayed similar behaviour, with a fall in growth level at the start of these competitions, followed by a rise, indicating the remarkable adaptability of pathogenic strain SH1000. The decrease in \textsl{S. aureus} density was related to the toxin level and growth rate of the respective species. 

Furthermore, as shown in Fig (\ref{Fig:f10d}), there was a significant link between the toxicity of the developed strain of \textsl{S. epidermidis} and the time required for \textsl{S. aureus} to adapt and mutate against these toxins. The most toxic species, TU3298, was able to inhibit SH1000 for a longer period.\\

\begin{figure}[H]
\centering
\begin{subfigure}{0.32\linewidth}
\centering
\caption{}
\label{Ln(B180_SH1000)}
\includegraphics[width=1\textwidth]{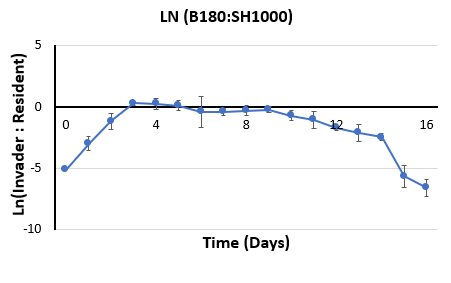}
\end{subfigure}
\hfill
\begin{subfigure}{0.32\linewidth}
\centering
\caption{}
\label{Ln(SH1000_B180)_1-1}
\includegraphics[width=1\textwidth]{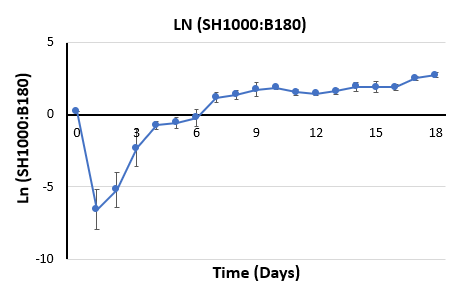}
\end{subfigure}
\hfill
\begin{subfigure}{0.32\linewidth}
\centering
\caption{}
\label{Ln(SH1000_B180)}
\includegraphics[width=1\textwidth]{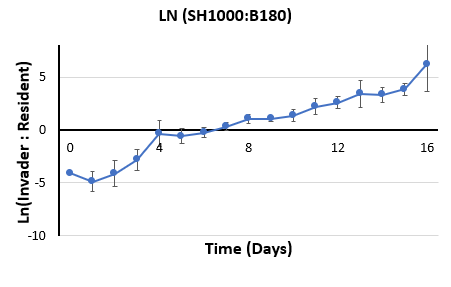}
\end{subfigure}
\\
\begin{subfigure}{0.32\linewidth}
\centering
\caption{}
\label{Ln(B155_SH1000)}
\includegraphics[width=1\textwidth]{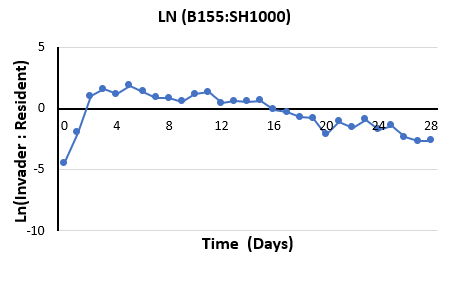}
\end{subfigure}
\hfill
\begin{subfigure}{0.32\linewidth}
\centering
\caption{}
\label{Ln(SH1000_B155)_1-1}
\includegraphics[width=1\textwidth]{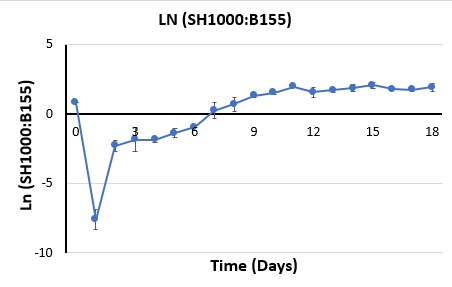}
\end{subfigure}
\hfill
\begin{subfigure}{0.32\linewidth}
\centering
\caption{}
\label{Ln(SH1000_B155)}
\includegraphics[width=1\textwidth]{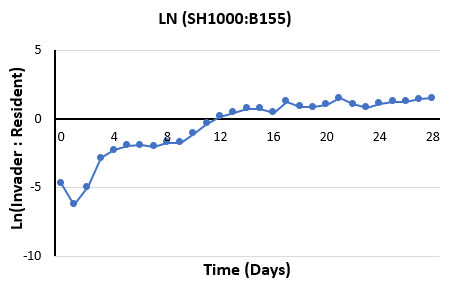}
\end{subfigure}
\\
\begin{subfigure}{0.32\linewidth}
\centering
\caption{}
\label{Ln(TU3298_SH1000)}
\includegraphics[width=1\textwidth]{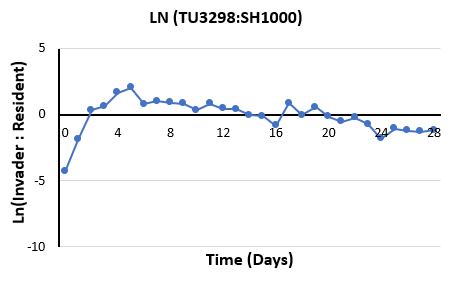}
\end{subfigure}
\hfill
\begin{subfigure}{0.32\linewidth}
\centering
\caption{}
\label{Ln(SH1000_TU3298)_1-1}
\includegraphics[width=1\textwidth]{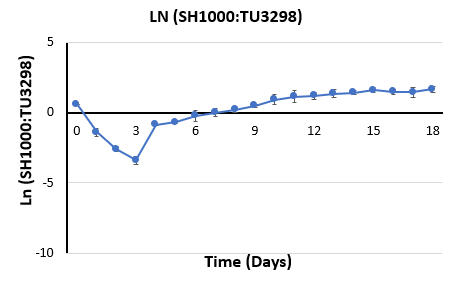}
\end{subfigure}
\hfill
\begin{subfigure}{0.32\linewidth}
\centering
\caption{}
\label{Ln(SH1000_TU3298)}
\includegraphics[width=1\textwidth]{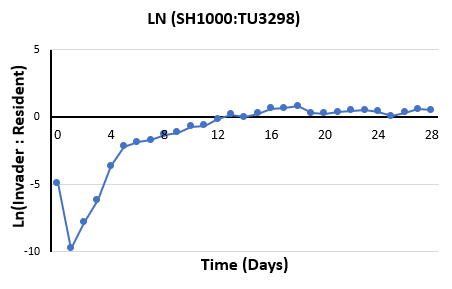}
\end{subfigure}
\caption[Plots of the natural log of the invader to resident ratio over time.]{\textbf{Plots of the natural log of the evolved population ratio over time when interactions were conducted under mixed conditions.}\,Panels (a), (d) and (g): show the natural log of the invader to resident ratio, (B180: SH1000), (B155: SH1000) and (TU3298: SH1000) respectively. Panels (b), (e) and (h): display the natural log of the interacting population ratios, (SH1000: B180), (SH1000: B155), and (SH1000: TU3298), in that order, when starting from equal initial concentrations. Panels (c), (f), and (i): represent the natural log of the invader to resident ratio when performing reciprocal invasions. Again, the $x$-axis is the time in days. Error bars represent the standard error of the mean (n = 3).}
\label{Mixed ratios}
\end{figure}
{\textbf{A negative association between the initial density of the \textsl{S. aureus} population and their ability to recover.}}

Unlike that found in \cite{A}, invasion of \textsl{S. aureus} into a toxin-producing \textsl{S. epidermidis} resident was positively frequency-dependent with the highest initial frequencies invading the fastest and lower initial frequencies becoming extinct. The interactions in this study shown in Fig (\ref{Fig:f10d}), revealed that the evolved \textsl{S. aureus} populations were able to recover faster when they started from lower initial frequencies (See figures \ref{C3}, \ref{C6}, and \ref{C9}). The evolved \textsl{S. aureus}, on the other hand, struggled when they started from higher initial frequencies (resident), as shown in figures (\ref{C1}), (\ref{C4}), and (\ref{C7}).

{\textbf{A positive association between the level of toxicity produced by \textsl{S. epidermidis} populations and their ability to persist.}}

In all performed interactions, no eliminations or complete displacements were observed. All evolved populations coexisted. However, S. epidermidis was more likely to persist at low frequencies, as seen in Fig (\ref{Fig:f10d}), and their chances of survival were positively associated with their level of toxicity.\\
{\textbf{The oscillations of the evolved population density were more evident when the interactions started from different initial concentrations.}}

As seen in Fig (\ref{Fig:f10d}), the invasion scenarios, when the evolved populations started from different initial concentrations rather than equal initial frequencies, the fluctuations in the density of evolving populations were more obvious and noticeable. Additionally, as seen in figures (\ref{C4}, \ref{C6}, \ref{C7}, and \ref{C9}), the production of toxins contributed favourably to this phenomenon.

{\textbf{\subsection{Mathematical Models and Simulations}}}
It is possible to expand on the spatially homogeneous \textsl{Lotka-Volterra} competition model to incorporate the diffusive terms of the respective species, $u$ and $v$. This creates the \eqref{Fig:num} system. As shown in Fig (\ref{environmental structure}), the interacted populations form a spot in the middle of the plate and expand symmetrically. Therefore, ideally, when modelling the interactions of microbial communities in Petri dishes, the Laplace operator in polar coordinates is used to express the space factor \eqref{system--2}.

\begin{equation}
\begin{cases} 
\displaystyle u_t = D_u \,u_{xx}+r_u\,u\,(1-u-b_1\,v),\\[8pt]
\displaystyle v_t= D_v \,v_{xx}+r_v\,v\,(1-v-b_2\,u).
\end{cases} 
\label{Fig:num}
\end{equation}
where $u=u(x,t)$ and $v=v(x,t)$ are concentrations of \textsl{\textsl{S. aureus}} and \textsl{\textsl{S. epidermidis}} strains. Here:
\begin{itemize}
\item  $x$ is the space variable, so $x \in [0, L]$, where $L$ is the length of the medium. (Size of Petri dish)
\item $t$ is the time variable, so $t \geq 0$. 
\item $r$ the growth rate of the strains.
\item $D$ diffusion coefficient.
\item $b$ effect that each strain has on the other.
\end{itemize}

\begin{equation}
\begin{cases} 
\displaystyle u_t = D_u \Big[u_{rr}+\frac{1}{r}\, u_r\Big]+r_u\,u\,(1-u-b_1\,v),\\[8pt]
\displaystyle v_t= D_v \Big[v_{rr}+\frac{1}{r}\, v_r\Big]+r_v\,v\,(1-v-b_2\,u).
\end{cases} 
\label{system--2}
\end{equation}

\noindent where $r$ is the radial distance. It should be noted that when scaling the space, larger $r$ would mean smaller diffusion coefficients $D$ and vice versa. Consequently, in any situation involving a small $D$, it is to be expected that any simulations produced by (\ref{system--2}) would be identical to simulations produced by (\ref{Fig:num}). As shown in Fig (\ref{polar_1}), after setting the time unit to $t=1$, which is equivalent to one day, the growth rate for the evolved populations and the diffusion coefficients are defined as estimated in \eqref{diffusion coefficients1}. Simulations obtained from (\ref{system--2}), (blue), coincide with the (\ref{Fig:num}) simulations, (red), when using the values found in the study presented in this thesis. 
\begin{figure}[H]
\centering
\begin{subfigure}{0.32\linewidth}
\centering
\caption{}
\label{polar_1}
\includegraphics[width=0.9\textwidth]{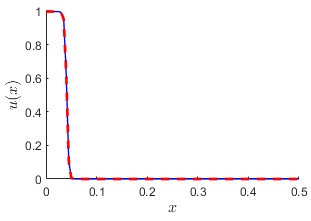}
\end{subfigure}
\hfill
\begin{subfigure}{0.32\linewidth}
\centering
\caption{}
\label{polar_2}
\includegraphics[width=0.9\textwidth]{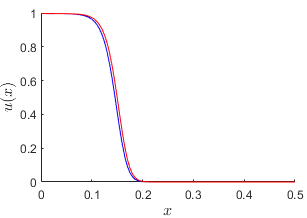}
\end{subfigure}
\hfill
\begin{subfigure}{0.32\linewidth}
\centering
\caption{}
\label{polar_3}
\includegraphics[width=0.9\textwidth]{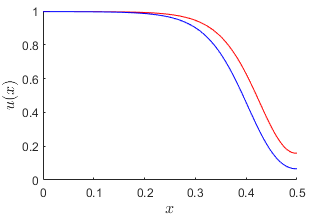}
\end{subfigure}
\caption{\textbf{A comparison of simulations produced by (\ref{Fig:num}) and (\ref{system--2}).}\,The profiles of $u$ are plotted in space at time ($T=10000\approx 12\,\text{hours}$), in the one-dimensional (red) and radially symmetric domains (blue). Panel (a): simulations obtained using the actual estimated diffusion coefficients in this study, $D_u=1.8\times\,10^{-6}$. Panel (b): $D_u=0.001$. Panel (c): $D_u=0.01 $. Parameters: $u_0=1$, $v_0=0.01$, $r_u=26.5296$, $r_v=28.613$, $b_1=0.1$ and $b_2=0.1$.}
\label{polar}
\end{figure}
Further investigations were conducted to see if this added term would make a difference. Thus, the space step was fixed to ensure the accuracy and the time step was fixed to satisfy the stability conditions of all the simulations, with and without the first derivative, and the diffusion coefficients were manipulated in both variables. As previously stated, gradually increasing the diffusion coefficients for the evolved populations is equivalent to decreasing $r$. It was observed that the simulations obtained from (\ref{system--2}) and (\ref{Fig:num}) would differ slightly, as shown in Fig (\ref{polar_2}), if the diffusion coefficients were significantly larger, (around $600$ times larger), in comparison to the actual estimated values in this study. When the diffusion coefficients were increased about $6000$ times more than the actuals, as seen in Fig (\ref{polar_3}), it was observed that the speed of the right travelling waves of the 1-D model in (\ref{Fig:num}), (red), was higher than the speed of the radially symmetric solutions, (blue). This delay was caused by the first derivative term in (\ref{system--2}).
%
The additional term in (\ref{system--2}), which includes the first derivatives, makes a difference if the diffusion coefficients are significant, as illustrated in Fig (\ref {polar_1}). Numerically distinguishing between the 1-D wave solution of (\ref{Fig:num}) and the corresponding front solution of (\ref{system--2}) poses a difficulty as they coincide when $D$ is small. Furthermore, as shown in Fig (\ref{diffusion coefficients}) and determined in (\ref{diffusion coefficients1}), when studying the diffusivities of the involved bacterial species, it is clear that all the bacterial species used diffuse weakly to the point that in many studies they are considered as non-motile species \cite{Carrol K}. Thus, it is considered that for the given $D's$ the solutions of (\ref{system--2}) will be well approximated by solutions of (\ref{Fig:num}). 
{\textbf{\subsubsection{Modelling Non-inhibitory Interactions}}}

The two-variable model simulations show a monotonic behaviour, meaning that the rate of change in population concentrations over time does not change the sign (i.e., if one of the competing populations is increasing, it is continuously increasing, and if it is decreasing it is continuously decreasing). Theoretically, to show that the two-variable model has a monotonic behaviour, the initial step was to define the expression $\frac{d}{dt}\big(ln(\displaystyle\frac{u}{v})\big)$, as follows:
$$
\displaystyle \frac{d}{dt}\Big(ln\Big(\displaystyle\frac{u}{v}\Big)\Big)=\displaystyle\frac{v}{u}\,\,.\,\, \displaystyle\frac{v.u^'-u.v^'}{v^2},
$$
from (\ref{Fig:num}), implanting the definition of $u_t$ and $v_t$ when assuming that $ D_u=D_v=0$, yields:
$$
\displaystyle \frac{d}{dt}\Big(ln\Big(\displaystyle\frac{u}{v}\Big)\Big)=\displaystyle\frac{v}{u}\,\,.\,\, \displaystyle\frac{v.r_u u(1-u-b_1 v)-u.r_v v(1-v-b_2 u)}{v^2}.
$$
For simplicity, as the growth rate for both competing populations are positive and comparable to each other, it was decided to set $r_u= r_v=1$, Hence:
$$
\displaystyle\frac{d}{dt}\Big(ln\Big(\displaystyle\frac{u}{v}\Big)\Big)=\Big[(b_2-1)u+(1-b_1)v\Big].
$$
Since all the competitions outcomes converge to the second equilibrium point $(u^*, v^*) = (1, 0)$, this equilibrium is stable if $b_1<1, b_2 > 1$. Hence:
\begin{equation}
\displaystyle\frac{d}{dt}\Big(ln\Big(\displaystyle\frac{u}{v}\Big)\Big)=a_1 u + a_2 v,
\label{always positive}
\end{equation}
where $a_1$ and $a_2$ are positive constants and $u,v \in [0,1]$. This means that rate of change in population concentrations, (SH1000: B180), over time is always positive, $\frac{d}{dt}\Big(ln\Big(\frac{u}{v}\Big)\Big)\geq0$, and this contradicts the findings presented in Fig (\ref{Ln(SH1000_B180)}), and Fig (\ref{Ln(SH1000_B180)_1-1}).

On the other hand, when defining $\frac{d}{dt}\big(ln(\frac{v}{u})\big)$, as follows:
$$
\displaystyle\frac{d}{dt}\Big(ln\Big(\displaystyle\frac{v}{u}\Big)\Big)=\displaystyle\frac{u}{v}\,\,.\,\, \displaystyle\frac{u.v^'-v.u^'}{u^2}.
$$
Following the previous steps, this expression was obtained: 
$$
\displaystyle\frac{d}{dt}\Big(ln\Big(\displaystyle\frac{v}{u}\Big)\Big)=\Big[(1-b_2)u+(b_1-1)v\Big].
$$
Again, applying the stability conditions gives:
\begin{equation}
\displaystyle\frac{d}{dt}\Big(ln\Big(\displaystyle\frac{v}{u}\Big)\Big)=-\Big[a_1 u + a_2 v\Big].
\label{always negative}
\end{equation}
This signifies that rate of change in population concentrations, (B180: SH1000), over time is always negative, and this contradicts the findings presented in Fig (\ref{Ln(B180_SH1000)}). From (\ref{always positive}) and (\ref{always negative}), it can be concluded that all possible simulations obtained from two-variable models that satisfy the stability conditions have monotonic behaviour. Such behaviour is inconsistent with the dynamics of interaction observed in the laboratory results. 

Thus, the two-variable model failed to simulate the illustrated experimental dynamics. Hence, the use of a two-variable model to simulate the experimental data was discarded.\\
{\textbf{Modelling The Dynamics of Interactions Between Two Populations With Adaptation}}

According to the experimental data shown in figures (\ref{Ln(B180_SH1000)}), (\ref{Ln(SH1000_B180)}) and (\ref{Ln(SH1000_B180)_1-1}), the evolved \textsl{S. aureus} exhibited a reduction in population size at the start of all interactions before recovering and becoming dominant. This reduction means that a large part of the interacted \textsl{S. aureus} population could not survive the competition and died. In contrast, a small part of the same population developed resistance against the opponent, which led to their success in achieving dominance. This dynamic of \textsl{S. aureus} population explains the non-monotonicity observed when plotting the natural logarithm of the competing population ratio. Mathematically, such a dynamic could be presented and generated if a third variable is added to represent the adapted part of the population. This part acts differently when interacting with \textsl{S. epidermidis}, as it is less sensitive to the effects of the opponent and competes more strongly to the point where it excludes its opponent. As a result, in the following section, a three-variable model is introduced. In this model, two variables represent the fractions of the \textsl{S. aureus} population known as susceptible, $u_s$, and adapted, $u_a$.

Two competing bacterial populations in a one-dimensional domain are considered. One species, the ‘susceptible’, has a fraction of cells capable of adapting to the fierceness of the competition, the ‘adapted’. Denoting the concentrations of susceptible \textsl{S. aureus}, adapted \textsl{S. aureus} and \textsl{S. epidermidis} at position $x$ and time $t$ by $u_s(x, t)$, $u_a(x, t)$ and $v_(x, t)$, this ecosystem is modelled via the equations:
\begin{equation}
\begin{cases} 
\displaystyle \frac{\partial u_s}{\partial t} = D_{u_s}\,\displaystyle\pd[2]{u_s}{x^2}+r_{u_{s}}\,u_s\,(1-u_s-b_{1}\,v-\,(1+\psi)\,u_a),\\[8pt]
\displaystyle\frac{\partial u_a}{\partial t} = D_{u_a}\,\displaystyle\pd[2]{u_a}{x^2}+r_{u_{a}}\,u_a\,(1-u_a-b_{2}\,v-\,(1+\psi)\,u_s),\\[8pt]
\displaystyle\frac{\partial v}{\partial t} = D_v\,\displaystyle\pd[2]{v}{x^2}+r_v\,v\,(1-v-b_{3}\,u_s-b_{4}\,u_a),\\
\end{cases} 
\label{system 2}
\end{equation}
where $D$, $r$, $b$ and $\psi$ terms are all positive constants. $D_{u_s}, D_{u_a}$ and $D_v$ represent the respective diffusion coefficient of each population, $r_{u_s}, r_{u_a}$ and $r_v$ represent the linear birth rates of each species. The $b$ terms measure the competitive effect of each population on the other, while $\psi\ll 1$ is a small value-added to the interaction coefficients between the susceptible and the adapted fractions of the \textsl{S. aureus} population to prevent obtaining zero as an eigenvalue when analysing the equilibrium points of this system.\\
Initial conditions:
$$
v(x,0) \displaystyle\begin{cases} 
1\,\,\,\,\,\,\,\,\,\,\,\,\,\,\,x\in[\frac{L}{2}-l,\frac{L}{2}+l],\\
0\,\,\,\,\,\,\,\,\,\,\,\,\,\,\,\text{otherwise},\\
\end{cases} 
$$
$$
\displaystyle u_s(x,0)+u_a(x,0)= 0.01\,\times\,v(x,0)\,\,\,\text{and}\,\,u_a/u_s=0.01.
$$
This can be presented the other way around when performing the mutual invasions.

Boundary conditions:
Again, zero–flux boundary conditions are imposed for all variables.\\
Similarly, as shown in two-variable model, in the absence of spatial variation, the system (\ref{system 2}) is considered as the three species food web model with the \textsl{Lotka-Volterra} type interaction between populations \cite{18}. There are six physically relevant (i.e., real and non-negative) stationary homogeneous solutions $(u_s, u_a, v)$=$ (u^*_s, u^*_a, v^*) $
\begin{itemize}
\item the trivial solution, $(u^*_s, u^*_a, v^*)=(0, 0, 0)$
\item $u_s$ excludes $u_a$ and $v$, $(u^*_s, u^*_a, v^*)=(1, 0, 0)$
\item $u_a$ excludes $u_s$ and $v$, $(u^*_s, u^*_a, v^*)=(0, 1, 0)$
\item $v$ excludes $u_s$ and $u_a$, $(u^*_s, u^*_a, v^*)=(0, 0, 1)$
\item $u_s$ and $v$ exclude $u_a$, $(u^*_s, u^*_a, v^*)=(\frac{1-b_1}{1-b_1\,b_3}, 0, \frac{1-b_3}{1-b_1\,b_3})$
\item $u_a$ and $v$ exclude $u_s$, $(u^*_s, u^*_a, v^*)=(0, \frac{1-b_2}{1-b_2\,b_4}, \frac{1-b_4}{1-b_2\,b_4})$
\end{itemize}
The two equilibriums that indicate co-existence between \textsl{S. aureus} and \textsl{S. epidermidis} are only considered if $u_s^*\geq 0$,\, $u_a^*\geq 0$\,and\, $v^*\geq 0$ are finite, in which case $ b_{1}\,.\, b_{3} \neq 1$ and $ b_{2}\,.\,b_{4} \neq 1$ respectively.
To determine the stability of the steady states, the Jacobian of communities is needed:\\
\begin{frame}

\resizebox{\linewidth}{!}{%
$\displaystyle
J=\left[ \begin{array}{rrrrrrrr} 
r_{1} \left(-b_{1}\,v -\mathit{u_s} -\mathit{u_a} +1\right)-r_{1} \mathit{u_s}  & -r_{1} \mathit{u_s}  & -r_{1} b_{1}\mathit{u_s}  \\
 -r_{2} \mathit{u_a}  & r_{2} \left(-b_{2} v -\mathit{u_s} -\mathit{u_a} +1\right)-r_{2} \mathit{u_a}  & -r_{2} b_{2} \mathit{u_a} \\
 -r_{3} b_{3}\,v  & -r_{3} b_{4} v  & r_{3} \left(-b_{3} \mathit{u_s} -\mathit{u_a} b_{4}-v +1\right)-r_{3} v\\
\end{array} \right].
$}
\end{frame}

The first steady state $(0, 0, 0)$, is unstable. By implementing this point into the Jacobian matrix, the positive eigenvalues $\lambda_1=r_1,\,\lambda_2=r_2$ and $\lambda_3=r_3$ are obtained.
$$
\displaystyle\left[\begin{array}{c} \lambda_{1}\\ \lambda_{2}\\\lambda_{3} \end{array}\right]_{(0, 0, 0)}=\displaystyle\left[\begin{array}{c} r_{1}\\ r_{2}\\ r_{3} \end{array}\right].
$$
Implementing the second equilibrium point, $(1, 0, 0)$, into the Jacobian matrix gives:
$$\left[\begin{array}{c} \lambda_{1}\\ \lambda_{2}\\\lambda_{3} \end{array}\right]_{(1, 0, 0)}=\left[\begin{array}{c} -r_{1}\\ -\frac{r_{2}}{10000}\\ -r_{3}\,\left(b_{3}-1\right) \end{array}\right].
$$  
The condition for this equilibrium to be stable is the following:
$$
(u_s^*, u_a^*, v^*)=(1, 0, 0) \,\,\,\, \text{is} \,\,\,\,\begin{cases} \mbox{stable}, & \mbox{if }\,\,\,\, b_3\,>1, \\ 
\mbox{unstable}, & \mbox{if }\,\,\,\,  b_3<1. \end{cases}
$$
The eigenvalues for the third equilibrium point, $(0, 1, 0)$, are: 
$$\left[\begin{array}{c} \lambda_{1}\\ \lambda_{2}\\\lambda_{3} \end{array}\right]_{(0, 1, 0)}=\left[\begin{array}{c} -r_{3}(b_{4}-1)\\ -r_{2}\\ -\frac{r_{1}}{10000} \end{array}\right].
$$  
The condition for this equilibrium to be stable is the following:
$$
(u_s^*, u_a^*, v^*)=(0, 1, 0) \,\,\,\, \text{is} \,\,\,\,\begin{cases} \mbox{stable}, & \mbox{if }\,\,\,\, b_4\,>1, \\ 
\mbox{unstable}, & \mbox{if }\,\,\,\,  b_4<1. \end{cases}
$$
When implementing the fourth equilibrium point, $(0, 0, 1)$, into the Jacobian matrix, the following eigenvalues were obtained:
$$\left[\begin{array}{c} \lambda_{1}\\ \lambda_{2}\\\lambda_{3} \end{array}\right]_{(0, 0, 1)}=\left[\begin{array}{c}
  -r_{1} (b_{1}-1) 
\\
 -r_{2} (b_2-1)
\\
-r_{3}
\end{array}\right].
$$  
The conditions for this equilibrium to be stable are the following:
$$
(u_s^*, u_a^*, v^*)=(0, 0, 1) \,\,\,\, \text{is} \,\,\,\,\begin{cases} \mbox{stable}, & \mbox{if }\,\,\,\, b_1, b_2\,>1, \\ 
\mbox{unstable}, & \mbox{if }\,\,\,\,  b_1, b_2\,<1. \end{cases}
$$

The eigenvalues for the equilibrium that indicates a co-existence between the susceptible \textsl{S. aureus} and \textsl{S. epidermidis} are:
$$\left[\begin{array}{c} \lambda_{1}\\ \lambda_{2}\\\lambda_{3} \end{array}\right]_{(\frac{1-b_1}{1-b_1\,b_3}, 0, \frac{1-b_3}{1-b_1\,b_3})}=\left[\begin{array}{c}
\frac{-r_{1} b_{1}-r_{3} b_{3}+r_{1}+r_{3}+\sqrt{r_{1}^{2} \left(-1+b_{1}\right)^{2}+4 \left(b_{1} b_{3}-\frac{1}{2}\right) \left(-1+b_{1}\right) r_{3} \left(b_{3}-1\right) r_{1}+r_{3}^{2} \left(b_{3}-1\right)^{2}}}{2 b_{1} b_{3}-2} 
\\
 \frac{-r_{1} b_{1}-r_{3} b_{3}+r_{1}+r_{3}-\sqrt{r_{1}^{2} \left(-1+b_{1}\right)^{2}+4 \left(b_{1} b_{3}-\frac{1}{2}\right) \left(-1+b_{1}\right) r_{3} \left(b_{3}-1\right) r_{1}+r_{3}^{2} \left(b_{3}-1\right)^{2}}}{2 b_{1} b_{3}-2} 
\\
 \frac{r_{2} \left(b_{3}-1\right) \left(b_{1}-b_{2}\right)}{b_{1} b_{3}-1} 
\end{array}\right].
$$
For simplicity, as the growth rate for both competing populations are positive and comparable to each other, it was decided to set $r_{u_s}= r_{u_a}= r_{v}=1$. Hence:
$$\left[\begin{array}{c} \lambda_{1}\\ \lambda_{2}\\\lambda_{3} \end{array}\right]_{(\frac{1-b_1}{1-b_1\,b_3}, 0, \frac{1-b_3}{1-b_1\,b_3})}=\left[\begin{array}{c}
-1 
\\
 \frac{\left(-1+b_{3}\right) \left(-1+b_{1}\right)}{b_{1} b_{3}-1} 
\\
 \frac{\left(-1+b_{3}\right) \left(b_{1}-b_{2}\right)}{b_{1} b_{3}-1} 
\end{array}\right].
$$
The conditions for this equilibrium to be stable are the following:
$$
(u_s^*, u_a^*, v^*)=(\frac{1-b_1}{1-b_1\,b_3}, 0, \frac{1-b_3}{1-b_1\,b_3})\,\,\,\, \text{is} \,\,\,\,\begin{cases} \mbox{stable}, & \mbox{if }\,\,\,\, b_1, b_3\,<1 \,\,\text{and}\,\,b_2>b_1,\\ 
\mbox{unstable}, & \mbox{if }\,\,\,\,  \text{otherwise}. \end{cases}
$$

$$\left[\begin{array}{c} \lambda_{1}\\ \lambda_{2}\\\lambda_{3} \end{array}\right]_{(0, \frac{1-b_2}{1-b_2\,b_4}, \frac{1-b_4}{1-b_2\,b_4})}=\left[\begin{array}{c}
\frac{-r_{2} b_{2}-r_{3} b_{4}+r_{2}+r_{3}+\sqrt{r_{2}^{2} \left(b_{2}-1\right)^{2}+4 \left(b_{2}-1\right) r_{3} \left(b_{2} b_{4}-\frac{1}{2}\right) \left(-1+b_{4}\right) r_{2}+r_{3}^{2} \left(-1+b_{4}\right)^{2}}}{2 b_{2} b_{4}-2} 
\\
 \frac{-r_{2} b_{2}-r_{3} b_{4}+r_{2}+r_{3}-\sqrt{r_{2}^{2} \left(b_{2}-1\right)^{2}+4 \left(b_{2}-1\right) r_{3} \left(b_{2} b_{4}-\frac{1}{2}\right) \left(-1+b_{4}\right) r_{2}+r_{3}^{2} \left(-1+b_{4}\right)^{2}}}{2 b_{2} b_{4}-2} 
\\
 -\frac{r_{1} \left(-1+b_{4}\right) \left(b_{1}-b_{2}\right)}{b_{2} b_{4}-1} 
\end{array}\right].
$$
Similarly, for simplicity, as the growth rate for both competing populations are positive and comparable to each other, it was decided to set $r_{u_s}= r_{u_a}= r_{v}=1$. Hence:
$$\left[\begin{array}{c} \lambda_{1}\\ \lambda_{2}\\\lambda_{3} \end{array}\right]_{(0, \frac{1-b_2}{1-b_2\,b_4}, \frac{1-b_4}{1-b_2\,b_4})}=\left[\begin{array}{c}
-1 
\\
 \frac{\left(-1+b_{4}\right) \left(-1+b_{2}\right)}{b_{2} b_{4}-1} 
\\
 \frac{\left(-1+b_{4}\right) \left(b_{2}-b_{1}\right)}{b_{2} b_{4}-1} 
\end{array}\right].
$$
The conditions for this equilibrium to be stable are the following:
$$
(u_s^*, u_a^*, v^*)=(0, \frac{1-b_2}{1-b_2\,b_4}, \frac{1-b_4}{1-b_2\,b_4})\,\,\,\, \text{is} \,\,\,\,\begin{cases} \mbox{stable}, & \mbox{if }\,\,\,\, b_2, b_4\,<1 \,\,\text{and}\,\,b_1>b_2,\\ 
\mbox{unstable}, & \mbox{if }\,\,\,\,  \text{otherwise}.\end{cases}
$$
From the previous equilibrium points, the only interest is in the third point, (0, 1, 0), where the adapted fraction of the \textsl{S. aureus} population, $u_a$, exclude the susceptible fraction, $u_s$, and \textsl{S. epidermidis} population $v$. Thus, the stability of this particular point is sought. By applying the stability condition $b_4 > 1$, better simulations are obtained that satisfy the obtained experimental data. 

According to the actual experimental data shown in Fig (\ref{Fig:f10d}), when isolates of \textsl{S. epidermidis} invaded the resident population of \textsl{S. aureus}, these invasions resulted in a decrease in the resident population density. However, the resident population recovered from the impact of these invasions and were able to increase production to reach population domination. This recovery started on the sixth day, when they reached the ratio (1: 1) with the \textsl{S. epidermidis} population and overcame their opponent afterwards. The extended model was able to capture these dynamics, and, as shown in Fig (\ref{Three-variable model simulation Ch3 a}), the evolved population intersected on day six. Furthermore, Fig (\ref{Three-variable model simulation Ch3 b}) displays the dynamic of interactions more clearly, where the increase of the invader, the red line, was at the expense of the collapse of the susceptible fraction of resident, blue line, resulting in the growth and emergence of resistance, (green line). The non-monotonicity appeared in the dynamics of interference when plotting the ratio between the evolved population, Fig (\ref{Three-variable model simulation Ch3 c}), which was mainly caused by the behaviour of the \textsl{S. aureus} population during the competition as it adapted after the majority of its population collapsed at the beginning of these invasions. The three-variable model successfully simulated such behaviour, (See Fig\,\ref{Three-variable model simulation Ch3 c}). 

\begin{figure}[H]
\centering
\begin{subfigure}{0.32\linewidth}
\centering
\caption{}
\label{Three-variable model simulation Ch3 a}
\includegraphics[width=0.9\textwidth]{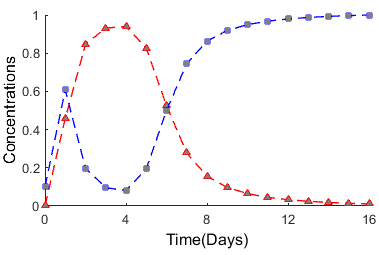}
\end{subfigure}
\hfill
\begin{subfigure}{0.32\linewidth}
\centering
\caption{}
\label{Three-variable model simulation Ch3 b}
\includegraphics[width=0.9\textwidth]{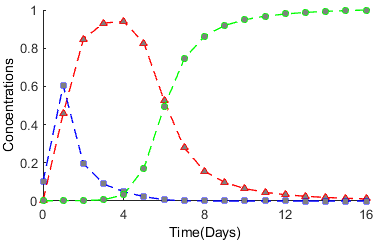}
\end{subfigure}
\hfill
\begin{subfigure}{0.32\linewidth}
\centering
\caption{}
\label{Three-variable model simulation Ch3 c}
\includegraphics[width=0.9\textwidth]{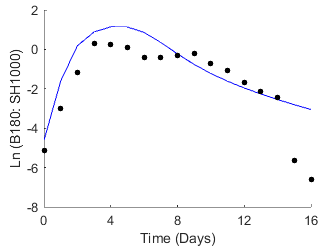}
\end{subfigure}
\caption{\textbf{Three-variable model simulation when low-toxin-producing isolates of \textsl{S. epidermidis} (B$180$) invade populations of \textsl{S. aureus} (SH$1000$) at frequencies of $0.01$.}\,Panel (a): simulations of interactions obtained from \eqref{system 2}; (red) is \textsl{S. epidermidis}; (blue) is \textsl{S. aureus}, $(u_s+u_a)$. Panel (b): demonstration of the three variable dynamics where (red) is \textsl{S. epidermidis}, (blue) is the susceptible fraction of \textsl{S. aureus} population, and (green) is the adapted part. Panel (c): the solid line indicates the simulation of the natural log of the invader to resident ratio, black dots represent the natural log of the invader to resident ratio from the actual data. In all panels, the $x$-axis is the time in days. The $y$-axis in panels (a) and (b) represents the relative concentrations of the evolved populations, and in panel (c) the natural log of the evolved population ratio. Parameters: $D_u=1.8\times\,10^{-6}$, $D_v=2\times\,10^{-5}$, $r_u=26.5296$, $r_v=29.7216$, $b_1=1.1$, $b_2=0.89$, $b_3=0.82$ and $b_4=1.45$.}
\label{Three-varaible model simulation CH3}
\end{figure}
Likewise, as shown in Fig (\ref{Three-varaible model simulation Ch33}), the three-variable model was able to simulate the other directions of the invasions when \textsl{S. epidermidis} populations became residents. Regardless of the oscillation that appeared in the behaviour of the \textsl{S. aureus} population during the interactions, and such behaviour was also observed in controls (See Fig \ref{control control}, dotted black lines), the obtained laboratory results showed that \textsl{S. aureus} was able to overcome its opponent by the fourth day when they intersected, (See Fig \ref{Ln(SH1000_B180)}).

The dynamic of interactions when performing this direction of invasion is illustrated evidently in Fig (\ref{Three-variable model simulation Ch33 b}), as the resident population of \textsl{S. epidermidis} (red line) restricted and inhibited the invasion by susceptible \textsl{S. aureus} (blue line). This led to the emergence of resistance by the invader population of \textsl{S. aureus} (green line).\\
{\textbf{Evolution from equal initial frequencies}}

As shown in Fig (\ref{C2}), and as previously noted, when the interaction between these two species, B180 and SH1000, started from equal initial concentration, it was observed that both populations tend to behave similarly when performing the invasions. 
\begin{figure}[H]
\centering
\begin{subfigure}{0.32\linewidth}
\centering
\caption{}
\label{Three-variable model simulation Ch33 a}
\includegraphics[width=0.9\textwidth]{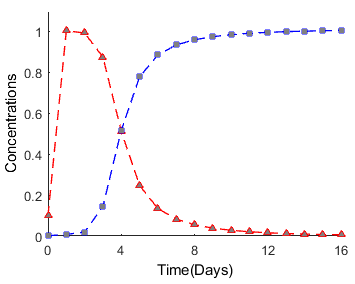}
\end{subfigure}
\hfill
\begin{subfigure}{0.32\linewidth}
\centering
\caption{}
\label{Three-variable model simulation Ch33 b}
\includegraphics[width=0.9\textwidth]{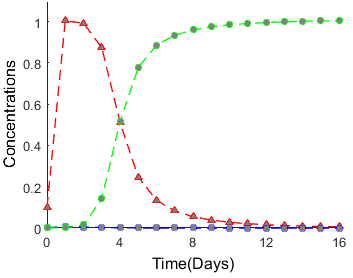}
\end{subfigure}
\hfill
\begin{subfigure}{0.32\linewidth}
\centering
\caption{}
\label{Three-variable model simulation Ch33 c}
\includegraphics[width=0.9\textwidth]{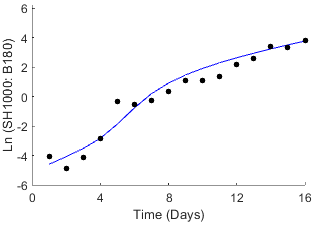}
\end{subfigure}
\caption{\textbf{Three-variable model simulation when low-toxin-producing isolates of \textsl{S. epidermidis} (B180) invaded by populations of \textsl{S. aureus} (SH1000) at frequencies of 0.01.}\,Panel (a): simulations of interactions obtained from \eqref{system 2}; (red) is \textsl{S. epidermidis}; (blue) is \textsl{S. aureus}, $(u_s+u_a)$. Panel (b): demonstration of the three variable dynamics where (red) is \textsl{S. epidermidis}; (blue) is the susceptible fraction of \textsl{S. aureus} population; and (green) is the adapted part. Panel (c): the solid line indicates the simulation of the natural log of the invader to resident ratio; black dots represent the natural log of the invader to resident ratio from the actual data. In all panels, the $x$-axis is the time in days. The $y$-axis in panels (a) and (b) represents the relative concentrations of the evolved populations, and in panel (c) the natural log of the evolved population ratio. Parameters: $D_u=1.8\times\,10^{-6}$, $D_v=2\times\,10^{-5}$, $r_u=26.5296$, $r_v=29.7216$, $b_1=1.1$, $b_2=0.89$, $b_3=0.82$ and $b_4=1.45$.}
\label{Three-varaible model simulation Ch33}
\end{figure}
The competing population of \textsl{S. epidermidis} diverged from the control experiment, where they were cultured independently, which indicates that the population was about to become extinct and vanish, and on the other hand the evolved \textsl{S. aureus} strain converged to its control experiment, from which it appears that their growth was no longer affected by the presence of the other competitor. However, the dynamic of the evolution seemed slower. As shown in Fig (\ref{Three-variable model simulation Ch333 a}), the competing population of \textsl{S. aureus} recovered and started to grow by day six, when they reached a (1: 1) ratio with \textsl{S. epidermidis} B180, (See Fig \ref{Three-variable model simulation Ch333 c}), and continued to evolve to become the majority of the developed populations.
\begin{figure}[H]
\centering
\begin{subfigure}{0.32\linewidth}
\centering
\caption{}
\label{Three-variable model simulation Ch333 a}
\includegraphics[width=0.9\textwidth]{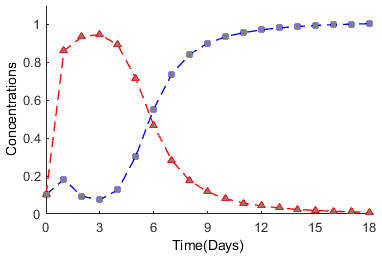}
\end{subfigure}
\hfill
\begin{subfigure}{0.32\linewidth}
\centering
\caption{}
\label{Three-variable model simulation Ch333 b}
\includegraphics[width=0.9\textwidth]{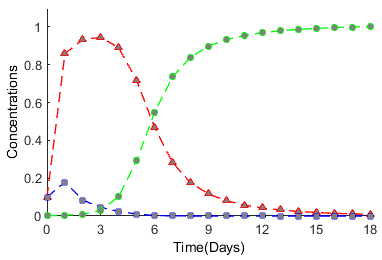}
\end{subfigure}
\hfill
\begin{subfigure}{0.32\linewidth}
\centering
\caption{}
\label{Three-variable model simulation Ch333 c}
\includegraphics[width=0.9\textwidth]{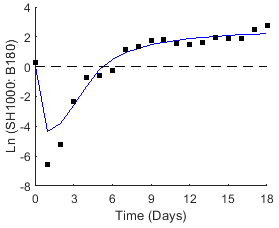}
\end{subfigure}
\caption{\textbf{Three-variable model simulation when low-toxin-producing isolates of \textsl{S. epidermidis} (B$180$) compete with populations of \textsl{S. aureus} (SH$1000$) at frequencies of $1: 1$.}\,Panel (a): simulations of interactions obtained from \eqref{system 2}; (red) is \textsl{S. epidermidis}; (blue) is \textsl{S. aureus}, $(u_s+u_a)$. Panel (b): demonstration of the three variable dynamics where (red) is \textsl{S. epidermidis}; (blue) is the susceptible fraction of \textsl{S. aureus} population; and (green) is the adapted part. Panel (c): the solid line indicates the simulation of the natural log of SH1000 to B180 ratio; black dots represent the natural log of this ratio from the actual data. In all panels, the $x$-axis is the time in days. The $y$-axis in panels (a) and (b) represents the relative concentrations of the evolved populations, and in panel (c) the natural log of the evolved population ratio. Parameters: $D_u=1.8\times\,10^{-6}$, $D_v=2\times\,10^{-5}$, $r_u=26.5296$, $r_v=29.7216$, $b_1=1.1$, $b_2=0.89$, $b_3=0.82$ and $b_4=1.45$.}
\label{Three-varaible model simulation Ch333}
\end{figure}
A detailed overview of the interactions is demonstrated in Fig (\ref{Three-variable model simulation Ch333 b}), where the evolved population of B180 (red line) inhibited and restricted the evolution of the susceptible fraction of \textsl{S. aureus} population (blue line) causing them to develop a resistance against their competitor (green line) and grow exponentially to form the major component of the population sample.

Considering the simulations shown in Fig (\ref{Three-varaible model simulation CH3}), Fig (\ref{Three-varaible model simulation Ch33}), and Fig (\ref{Three-varaible model simulation Ch333}), it can be concluded that the three-variable model could generate a demonstration of the interaction dynamics. However, this model was not able to create the oscillation observed in the experimental data.
{\textbf{\subsubsection{Modelling Inhibitory Interactions}}}
The aim of this study is to model the interactions involving toxin production, where one of the competing populations, \textsl{S. epidermidis}, produces toxin at a rate of $f_1$, while the other competing population, \textsl{S. aureus}, is well-known for its ability to adapt and mutate against these toxins. The concentrations of susceptible \textsl{S. aureus}, adapted {S. aureus}, producer \textsl{S. epidermidis}, and toxin at position $x$ and time $t$  are defined as $u_s (x, t)$, $u_a (x, t)$, $v (x, t)$, and $T (x, t)$.

Thus, this ecosystem is modelled via the equations:
\begin{equation}
\begin{cases} 
\displaystyle \frac{\partial u_s}{\partial t} = D\displaystyle\pd[2]{u_s}{x^2}+r_{u_{s}}\,u_s\,(1-u_s-b_{1}\,v-\,(1+\psi)\,u_a-p_{1}\,T),\\[8pt]
\displaystyle\frac{\partial u_a}{\partial t} = D\displaystyle\pd[2]{u_a}{x^2}+r_{u_{a}}\,u_a\,(1-u_a-b_{2}\,v-\,(1+\psi)\,u_s-p_{2}\,T),\\[8pt]
\displaystyle\frac{\partial v}{\partial t} = D\displaystyle\pd[2]{v}{x^2}+r_v\,v\,(1-v-b_{3}\,u_s-b_{4}\,u_a),\\[8pt]
\displaystyle\frac{\partial T}{\partial t} = D_T\displaystyle\pd[2]{T}{x^2}+f_1\,v\,-f_2\,T,\\
\end{cases} 
\label{system3}
\end{equation}
where $r$'s represent the growth rates and $b$'s are the competition coefficients,\,$p$: inhibition coefficient, $f_1$: inhibitor production rate, $f_2$: inhibitor degradation rate. While $\psi\ll 1$ is a small value-added to the interaction coefficients between the susceptible and the adapted fractions of the \textsl{S. aureus} population to prevent obtaining zero as an eigenvalue when analysing the equilibrium points of this system.\\
Initial conditions:
$$
v(x,0)\begin{cases} 
1\,\,\,\,\,\,\,\,\,\,\,\,\,\,\,x\in[\frac{L}{2}-l,\frac{L}{2}+l],\\[8pt]
0\,\,\,\,\,\,\,\,\,\,\,\,\,\,\,\text{otherwise},\\
\end{cases} 
$$
$$
\displaystyle u_s(x,0)+u_a(x,0)= a\,.\,v(x,0)\,\,\text{where}\,a\,=\,0.01\,\, \text{and}\,\,u_a/u_s=0.01\, \text{and}\, T(x,0)=0.
$$
This is also valid contrariwise when performing the mutual invasions.\\
Boundary conditions:
Again, zero–flux boundary conditions are imposed for all variables.\\
This model (\ref{system3}) represents the spatially extended \textsl{Lotka-Volterra} model when two species compete in one dimension. 
\begin{figure}[H]
\centering
\begin{subfigure}{0.32\linewidth}
\centering
\caption{}
\label{Four-variable model simulation CH3 a}
\includegraphics[width=0.9\textwidth]{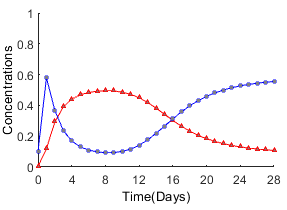}
\end{subfigure}
\hfill
\begin{subfigure}{0.32\linewidth}
\centering
\caption{}
\label{Four-variable model simulation CH3 b}
\includegraphics[width=0.9\textwidth]{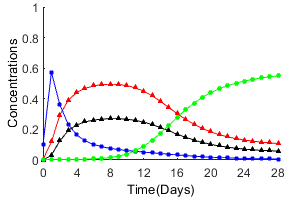}
\end{subfigure}
\hfill
\begin{subfigure}{0.32\linewidth}
\centering
\caption{}
\label{Four-variable model simulation CH3 c}
\includegraphics[width=0.9\textwidth]{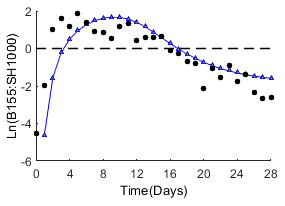}
\end{subfigure}
\caption{\textbf{Four-variable model simulation when toxin-producing isolates of \textsl{S. epidermidis} (B155) invade populations of \textsl{S. aureus} (SH1000) at frequencies of 0.01: 1.}\,Panel (a): simulations of interactions obtained from \eqref{system3}; (red) is \textsl{S. epidermidis}; (blue) is \textsl{S. aureus}, $(u_s+u_a)$.  Panel (b): demonstration of the four variable dynamics where (red) is \textsl{S. epidermidis}; (blue) is the susceptible fraction of \textsl{S. aureus} population; (green) is the adapted part; and (black) is the toxin. Panel (c): the solid line indicates the simulation of the natural log of the invader to resident ratio; black dots represent the natural log of the invader to resident ratio from the actual data. In all panels, the $x$-axis is the time in days. The $y$-axis in panels (a) and (b) represents the relative concentrations of the evolved populations and toxins, and in panel (c) the natural log of the evolved population ratio. Parameters: $D_u=1.8\times\,10^{-6}$, $D_v=5\times\,10^{-6}$, $r_u=26.5296$, $r_v=22.3344$, $p_1=1.0091$, $p_2=0.9755$, $f_1=1$, $f_2=0.18$, $b_1=1.1$, $b_2=0.9$, $b_3=0.65$ and $b_4=0.96$.}
\label{Four-varaible model simulation CH3}
\end{figure}
The modifications were made to account for the production of toxins by one species to inhibit the other and the ability of the inhibited population to mutate against these toxins. This model has more parameters, as stated and defined previously, in compression with the other models. However, all the added parameters which represent inhibition coefficients, production, and decaying rate of the toxins can be estimated and defined by using the equation (\ref{f_1 and f_2 equation}).

As illustrated in Fig (\ref{f_1 and f_2 curves}), there is a positive association between the toxin production rate and the decaying rate. The same correlation is observed between the inhibition coefficients and the decaying rate. 

\begin{figure}[H]
\centering
\begin{subfigure}{0.32\linewidth}
\centering
\caption{}
\label{Four-variable model simulation CH33 a}
\includegraphics[width=0.9\textwidth]{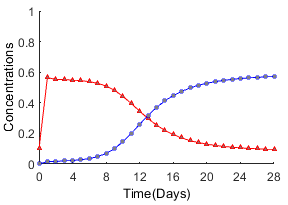}
\end{subfigure}
\hfill
\begin{subfigure}{0.32\linewidth}
\centering
\caption{}
\label{Four-variable model simulation CH33 b}
\includegraphics[width=0.9\textwidth]{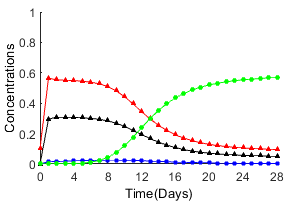}
\end{subfigure}
\hfill
\begin{subfigure}{0.32\linewidth}
\centering
\caption{}
\label{Four-variable model simulation CH33 c}
\includegraphics[width=0.9\textwidth]{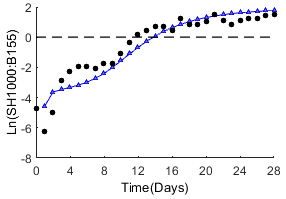}
\end{subfigure}
\caption{\textbf{Four-variable model simulation when toxin-producing isolates of \textsl{S. epidermidis} (B$155$) invaded by populations of \textsl{S. aureus} (SH$1000$) at frequencies of $0.01: 1$.}\,Panel (a): simulations of interactions obtained from \eqref{system3}; (red) is \textsl{S. epidermidis}; (blue) is \textsl{S. aureus}, $(u_s+u_a)$.  Panel (b): demonstration of the four variable dynamics where (red) is \textsl{S. epidermidis}; (blue) is the susceptible fraction of \textsl{S. aureus} population; (green) is the adapted part; and (black) is the toxin. Panel (c): the solid line indicates the simulation of the natural log of the invader to resident ratio; black dots represent the natural log of the invader to resident ratio from the actual data. In all panels, the $x$-axis is the time in days. The $y$-axis in panels (a) and (b) represents the relative concentrations of the evolved populations and toxins, and in panel (c) the natural log of the evolved population ratio. Parameters: $D_u=1.8\times\,10^{-6}$, $D_v=5\times\,10^{-6}$, $r_u=26.5296$, $r_v=22.3344$, $p_1=1.0091$, $p_2=0.9755$, $f_1=1$, $f_2=0.18$, $b_1=1.1$, $b_2=0.9$, $b_3=0.65$ and $b_4=0.96$.}
\label{Four-varaible model simulation CH33}
\end{figure}
When different values of $f_2$ were tested in the model (\ref{system3}), it was discovered that higher values of the declining rate, $f_2$, imply higher values of the production rate, $f_1$, as the results of these \textsl{S. aureus} populations would not have a chance to successfully invade the \textsl{S. epidermidis} populations, which contradicts the experimental findings of this study. See figures (\ref{C4}), (\ref{C5}), and (\ref{C6}).
\begin{figure}[H]
\centering
\begin{subfigure}{0.32\linewidth}
\centering
\caption{}
\label{Four-variable model simulation CH333 a}
\includegraphics[width=0.9\textwidth]{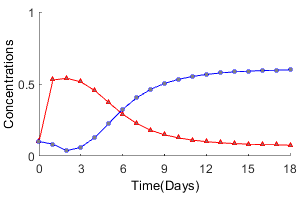}
\end{subfigure}
\hfill
\begin{subfigure}{0.32\linewidth}
\centering
\caption{}
\label{Four-variable model simulation CH333 b}
\includegraphics[width=0.9\textwidth]{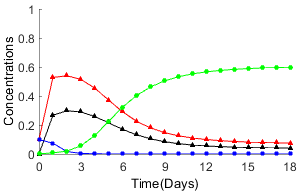}
\end{subfigure}
\hfill
\begin{subfigure}{0.32\linewidth}
\centering
\caption{}
\label{Four-variable model simulation CH333 c}
\includegraphics[width=0.9\textwidth]{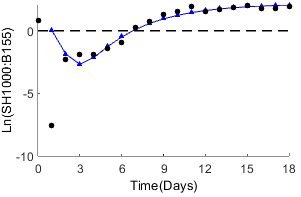}
\end{subfigure}
\caption{\textbf{Four-variable model simulation when toxin-producing isolates of \textsl{S. epidermidis} (B$155$) compete with populations of \textsl{S. aureus} (SH$1000$) at initial frequencies of $1: 1$.}\,Panel (a): simulations of interactions obtained from \eqref{system3}; (red) is \textsl{S. epidermidis}; (blue) is \textsl{S. aureus}, $(u_s+u_a)$.  Panel (b): demonstration of the four variable dynamics where (red) is \textsl{S. epidermidis}; (blue) is the susceptible fraction of \textsl{S. aureus} population; (green) is the adapted part; and (black) is the toxin. Panel (c): the solid line indicates the simulation of the natural log of SH1000 to B155 ratio; black dots represent the natural log of this ratio from the actual data. In all panels, the $x$-axis is the time in days. The $y$-axis in panels (a) and (b) represents the relative concentrations of the evolved populations and toxins, panel (c) the natural log of the evolved population ratio. Parameters: $D_u=1.8\times\,10^{-6}$, $D_v=5\times\,10^{-6}$, $r_u=26.5296$, $r_v=22.3344$, $p_1=1.0091$, $p_2=0.9755$, $f_1=1$, $f_2=0.18$, $b_1=1.1$, $b_2=0.9$, $b_3=0.65$ and $b_4=0.96$.}
\label{Four-varaible model simulation CH333}
\end{figure}
Furthermore, small values of $f_2$ mean that \textsl{S. aureus} will invade rapidly. Hence, it is necessary to choose the proper value that will satisfy the experimental data in this study. Thus, when the production rate is fixed at $f_1 = 1$, selecting the appropriate value of $f_2$ gives the corresponding values for $p_1$ and $p_2$, which are the inhibition coefficients of the toxins produced by B155 on the susceptible and adapted fractions of \textsl{S. aureus}, respectively. Four variable model simulations were fitted to the actual experimental data by using the least square method. It was possible to obtain the best values, with the minimum error, for the interaction coefficients as well as the inhibitory reduction rate after defining the inhibition coefficients $p_1$ and $p_2$ with respect to $f_2$, as presented in equation (\ref{f_1 and f_2 equation}).

\begin{figure}[H]
\centering
\begin{subfigure}{0.32\linewidth}
\centering
\caption{}
\label{Four-variable model simulation Ch3 a}
\includegraphics[width=0.9\textwidth]{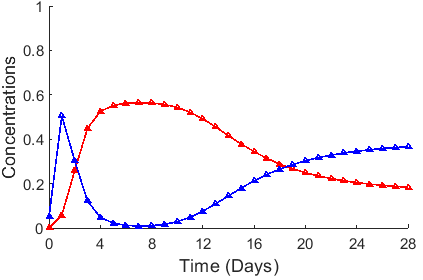}
\end{subfigure}
\hfill
\begin{subfigure}{0.32\linewidth}
\centering
\caption{}
\label{Four-variable model simulation Ch3 b}
\includegraphics[width=0.9\textwidth]{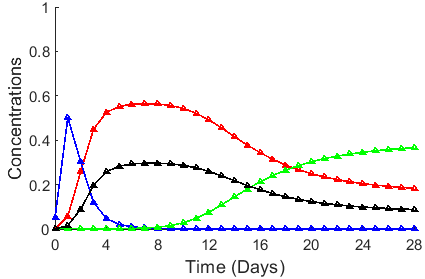}
\end{subfigure}
\hfill
\begin{subfigure}{0.32\linewidth}
\centering
\caption{}
\label{Four-variable model simulation Ch3 c}
\includegraphics[width=0.9\textwidth]{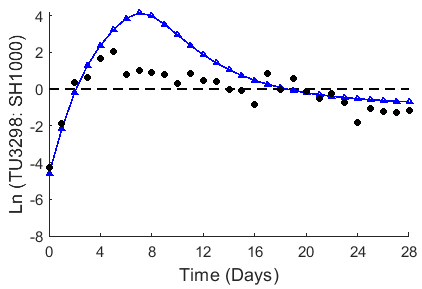}
\end{subfigure}
\caption{\textbf{Four-variable model simulation when toxin-producing isolates of \textsl{S. epidermidis} (TU$3298$) invade populations of \textsl{S. aureus} (SH$1000$) at frequencies of $0.01: 1$.}\,Panel (a): simulations of interactions obtained from \eqref{system3}; (red) is \textsl{S. epidermidis}; (blue) is \textsl{S. aureus}, $(u_s+u_a)$.  Panel (b): demonstration of the four variable dynamics where (red) is \textsl{S. epidermidis}; (blue) is the susceptible fraction of \textsl{S. aureus} population; (green) is the adapted part; and (black) is the toxin. Panel (c): the solid line indicates the simulation of the natural log of the invader to resident ratio; black dots represent the natural log of the invader to resident ratio from the actual data. In all panels, the $x$-axis is the time in days. The $y$-axis in panels (a) and (b) represents the relative concentrations of the evolved populations and toxins, and in panel (c) the natural log of the evolved population ratio. Parameters: $D_u=1.8\times\,10^{-6}$, $D_v=5\times\,10^{-5}$, $r_u=26.5296$, $r_v=28.613$, $p_1=0.7746$, $p_2=0.7623$, $f_1=1$, $f_2=0.24$, $b_1=0.9$, $b_2=0.84$, $b_3=0.9$ and $b_4=1.09$.}
\label{Four-varaible model simulation Ch3}
\end{figure}
Regardless of the oscillations observed in the dynamics of interactions when starting from different initial concentrations, figures (\ref{C4}) and (\ref{C6}), the four-variable model successfully simulated the experimental data that reflect interactions between toxin-producing \textsl{S. epidermidis} and \textsl{S. aureus} populations in terms of the intersection point between the competing populations as well as the final state of the evolved species, as seen in Fig (\ref{Four-varaible model simulation CH3}). 

When populations of the toxin-producing strain B155 were introduced to the resident population of \textsl{S. aureus}, SH1000, as a result the density of the resident population decreased and struggled until it was able to develop a resistance mechanism and raise its concentration back to the ratio of (1: 1) with its opponent by day $16 - 17$. Similarly, when the invasion was carried out in the opposite direction, the resident population of B155 restricted and inhibited the invasions by \textsl{S. aureus} up to day 8, after which the invader \textsl{S. aureus} mutated against the inhibitory produced by B155 and began to recover to the point where it coincided with B155 on day 12 and eventually dominated the interactions.

The model simulations presented in figures (\ref{Four-varaible model simulation CH3}), (\ref{Four-varaible model simulation CH33}), and (\ref{Four-varaible model simulation CH333}) captured these sort of behaviours when simulating the invasions. When using the four-variable model to simulate the interaction outcomes between B155 and SH1000, all parameters were fixed regardless of the initial concentrations of the competing populations.  

Likewise, when using a four-variable model to produce simulations of the interactions between populations of \textsl{S. aureus} and \textsl{S. epidermidis}, TU3298, the production, decaying and inhibition coefficients were fixed regardless of the initial concentrations of the competing populations.  

According to Fig (\ref{inhibition assay - before 1}), \textsl{S. epidermidis} strain TU3298 produced a larger inhibition zone in comparison with the other involved strains. However, according to \cite{19}, there are many technical factors influencing the size of the zone in the disc diffusion method. For instance, the density of inoculum, the inhibition zones will be larger if the inoculum is too light, even if the organism’s sensitivity remains unchanged. Relatively resistant strains can be considered as susceptible strains. If the inoculum is too heavy, the zone size is reduced, and susceptible strains may be reported as resistant.  
\begin{figure}[H]
\centering
\begin{subfigure}{0.32\linewidth}
\centering
\caption{}
\label{Four-variable model simulation Ch33 a}
\includegraphics[width=0.9\textwidth]{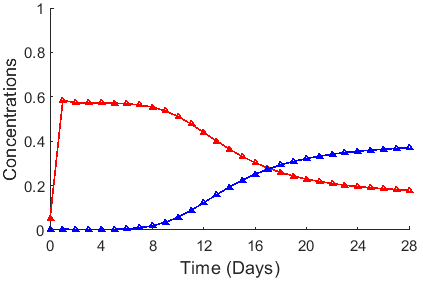}
\end{subfigure}
\hfill
\begin{subfigure}{0.32\linewidth}
\centering
\caption{}
\label{Four-variable model simulation Ch33 b}
\includegraphics[width=0.9\textwidth]{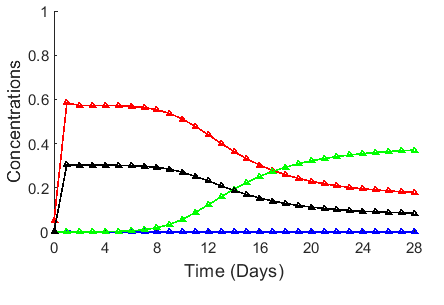}
\end{subfigure}
\hfill
\begin{subfigure}{0.32\linewidth}
\centering
\caption{}
\label{Four-variable model simulation Ch33 c}
\includegraphics[width=0.9\textwidth]{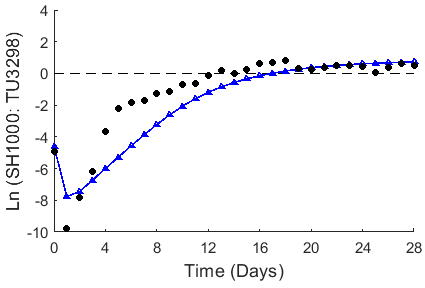}
\end{subfigure}
\caption{\textbf{Four-variable model simulation when toxin-producing isolates of \textsl{S. epidermidis} (TU$3298$) invaded by populations of \textsl{S. aureus} (SH$1000$) at frequencies of $0.01: 1$.}\,Panel (a): simulations of interactions obtained from \eqref{system3}; (red) is \textsl{S. epidermidis}; (blue) is \textsl{S. aureus}, $(u_s+u_a)$.  Panel (b): demonstration of the four variable dynamics where (red) is \textsl{S. epidermidis}; (blue) is the susceptible fraction of \textsl{S. aureus} population; (green) is the adapted part; and (black) is the toxin. Panel (c): the solid line indicates the simulation of the natural log of the invader to resident ratio; black dots represent the natural log of the invader to resident ratio from the actual data. In all panels, the $x$-axis is the time in days. The $y$-axis in panels (a) and (b) represents the relative concentrations of the evolved populations and toxins, and in panel (c) the natural log of the evolved population ratio. Parameters: $D_u=1.8\times\,10^{-6}$, $D_v=5\times\,10^{-5}$, $r_u=26.5296$, $r_v=28.613$, $p_1=0.7746$, $p_2=0.7623$, $f_1=1$, $f_2=0.24$, $b_1=0.9$, $b_2=0.84$, $b_3=0.9$ and $b_4=1.09$.}
\label{Four-varaible model simulation Ch33}
\end{figure}
Furthermore, the production rate of the producer can positively contribute to the size of the inhibition zone. As stated in \cite{Inhibition}, a growth inhibition experiment evaluates one bacterium’s ability to inhibit the growth of another by producing antimicrobial chemicals or competing for resources. Thus, the clear zone around the spot possibly signifies that it does not contain food, as it is consumed by the producer, and that it is no longer suitable for the growth of the other competing bacteria. 

As demonstrated in Table (\ref{tab:doubling time}), \textsl{S. epidermidis} strain TU3298 has a higher growth rate than its opponent, \textsl{S. aureus} SH1000 and in the model (\ref{system3}), the production rate of the toxins is positively associated with the growth rate of the producer. As it was decided to set the toxin production rate to $f_1 = 1$, it is necessary to define the minimum inhibitory coefficients $p_1$ and $p_2$ to allow \textsl{S. aureus} to invade and satisfy the experimental findings, as shown in figures (\ref{Four-varaible model simulation Ch3}), (\ref{Four-varaible model simulation Ch33}) and (\ref{Four-varaible model simulation Ch333}).
\begin{figure}[H]
\centering
\begin{subfigure}{0.32\linewidth}
\centering
\caption{}
\label{Four-variable model simulation Ch333 a}
\includegraphics[width=0.9\textwidth]{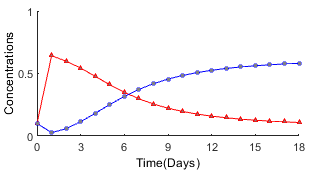}
\end{subfigure}
\hfill
\begin{subfigure}{0.32\linewidth}
\centering
\caption{}
\label{Four-variable model simulation Ch333 b}
\includegraphics[width=0.9\textwidth]{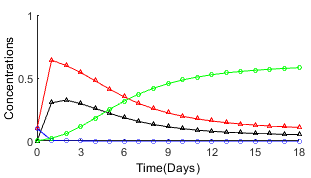}
\end{subfigure}
\hfill
\begin{subfigure}{0.32\linewidth}
\centering
\caption{}
\label{Four-variable model simulation Ch333 c}
\includegraphics[width=0.9\textwidth]{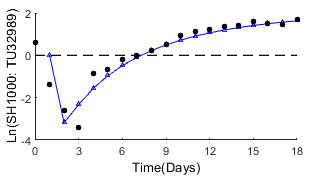}
\end{subfigure}
\caption{\textbf{Four-variable model simulation when toxin-producing isolates of \textsl{S. epidermidis} (TU$3298$) compete with populations of \textsl{S. aureus} (SH$1000$) at initial frequencies of $1: 1$.}\,Panel (a): simulations of interactions obtained from \eqref{system3}; (red) is \textsl{S. epidermidis}; (blue) is \textsl{S. aureus}, $(u_s+u_a)$.  Panel (b): demonstration of the four variable dynamics where (red) is \textsl{S. epidermidis}; (blue) is the susceptible fraction of \textsl{S. aureus} population; (green) is the adapted part; and (black) is the toxin. Panel (c): the solid line indicates the simulation of the natural log of SH1000 to TU3298 ratio; black dots represent the natural log of this ratio from the actual data. In all panels, the $x$-axis is the time in days. The $y$-axis in panels (a) and (b) represents the relative concentrations of the evolved populations and toxins, and in panel (c) the natural log of the evolved population ratio. Parameters: $D_u=1.8\times\,10^{-6}$, $D_v=5\times\,10^{-5}$, $r_u=26.5296$, $r_v=28.613$, $p_1=0.7746$, $p_2=0.7623$, $f_1=1$, $f_2=0.24$, $b_1=0.9$, $b_2=0.84$, $b_3=0.9$ and $b_4=1.09$.}
\label{Four-varaible model simulation Ch333}
\end{figure}
The values of the coefficients $p_1$ and $p_2$, the inhibiting factors, and the corresponding decay rate $f_2$ that satisfy the observed behaviour of the evolved \textsl{S. aureus} are determined using the fitting technique.

The best fit to the experimental data with the minimum error was obtained when $p_1=0.7746$, $p_2=0.7623$ and $f_2=0.24$ and the ratio between these factors were defined and illustrated in equation (\ref{f_1 and f_2 equation}) and plotted in Fig (\ref{f_1 and f_2 curves}).
The consequences of choosing slightly higher inhibition coefficients, that fit the ratio presented in Fig (\ref{f_1 and f_2 curves}), is an acceleration of the interference mechanism because the toxin-producing population will inhibit and restrict the susceptible fraction of the population faster. According to the model produced in this study, the decline of the susceptible fraction leads to the emergence of the resistance population. Thus, the whole mechanism will be accelerated. On the contrary, choosing slightly lower inhibition coefficients would delay the dynamics of interaction. For the same purposes, a lower inhibition coefficient allows the susceptible fraction of the population to sustain for a longer period which leads to a delay in the emergence of resistance. Illustrations of the effect of different values of $p_1$ and $p_2$ will be shown in the appendix Fig (\ref{different_fs}).

{\textbf{\section{Mathematical Investigation of Resistance Evolution}}}

This section aims to test the 3-4 variable model hypotheses generated as set out in the previous sections. According to the interaction outcomes, regardless of the initial concentrations of the evolved populations, \textsl{S. aureus} was able to dominate in every single interaction, meaning that \textsl{S. aureus} populations had developed resistance against the toxins produced by \textsl{S. epidermidis}, which explains the decline in their frequencies. These findings were also confirmed by results presented in section (3.2.3), as most of the evolved populations of \textsl{S. aureus} showed no inhibition zones when spread over the evolved populations of \textsl{S. epidermidis}.

However, some replicates showed smaller inhibition zones against the toxins produced by B155 and TU3298
\begin{figure}[H]
\centering
\begin{subfigure}{0.9\linewidth}
\centering
\caption{}
\label{testB180}
\includegraphics[width=0.7\textwidth]{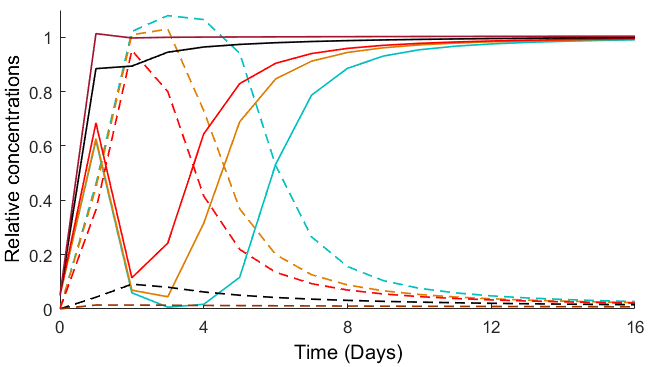}
\end{subfigure}
\\
\begin{subfigure}{0.9\linewidth}
\centering
\caption{}
\label{testB155}
\includegraphics[width=0.7\textwidth]{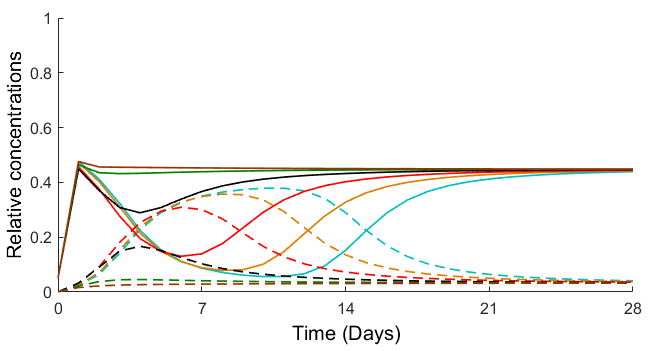}
\end{subfigure}
\\
\begin{subfigure}{0.9\linewidth}
\centering
\caption{}
\label{testTU3298}
\includegraphics[width=0.7\textwidth]{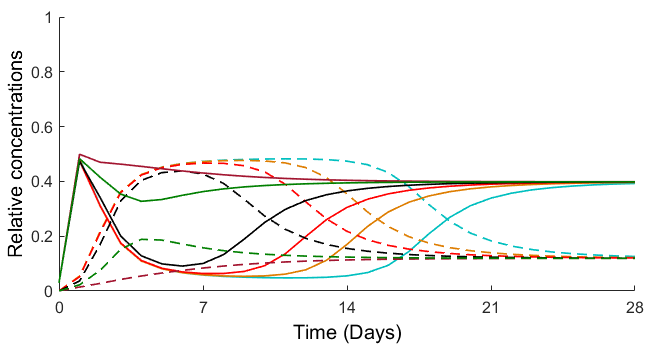}
\end{subfigure}
\caption{\textbf{Results of numerical experiments conducted to evaluate the modelling predictions and estimate the proportion of the adapted \textsl{S. aureus} SH1000.}\,Panel (a): Different simulations were generated from the three-variable model (\ref{system 2}), representing the interactions between \textsl{S. epidermidis} B180 (invader) and \textsl{S. aureus} SH1000 (resident) when the ratios of adaptive ($u_a$) and susceptible ($u_s$) \textsl{S. aureus} fractures were varied. Panels (b) and (c): different simulations were generated from the four-variable model (\ref{system3}), representing the interactions between \textsl{S. epidermidis} B155 (invader) and \textsl{S. aureus} SH1000 (resident), and between \textsl{S. epidermidis} TU3298 (invader) and \textsl{S. aureus} SH1000 (resident), when the ratios of adaptive ($u_a$) and susceptible ($u_s$) \textsl{S. aureus} fractures were varied. Dashed lines represent \textsl{S. epidermidis}; solid lines represent \textsl{S. aureus} ($u_s + u_a$). Different colours of lines indicate different ratios of the adapted fractures of SH1000, 70\% dark red, 60\% dark green, 50\% black, 30\% red, 10\% orange, and 1\% (initial assumption) light blue lines. The $x$-axis represents the time in days, and the $y$-axis is the relative concentrations of the evolved populations. The parameters used in panels (a), (b), and (c) were the same as the parameters in figures (\ref{Three-varaible model simulation CH3}), (\ref{Four-varaible model simulation CH3}), and (\ref{Four-varaible model simulation Ch3}), respectively.}
\label{Numerical experiments}
\end{figure}
compared to the initial inhibition zones (before interactions), indicating that the evolved isolates had not yet entirely mutated against these toxins and that isolates still contained a sensitive fraction of populations. 

Since the size of post-interaction inhibition zones is positively correlated with the level of toxicity of the evolved \textsl{S. epidermidis}, the author of this thesis considers that it is only a matter of time before the whole developed population of \textsl{S. aureus} becomes resistant to these toxins. It is also worth noting that all evolved strains of \textsl{S. epidermidis} maintain their existence at lower frequencies than \textsl{S. aureus}, and the chances of their survival are positively correlated with their level of toxicity. 

Thus, the final ratio between the evolved populations differs, \textsl{S. aureus} grew approximately ten times larger than B180, almost five times larger than B155, and approximately three times larger than TU3298. The three and four-variable model predictions could be tested in silico by changing the initial conditions, which assumed that the adapted fraction of \textsl{S. aureus} represented only 1\% of the population. The proportions of the adapted fraction of \textsl{S. aureus} SH1000 were gradually increased. Thus, a family of population curves (shown in Fig \ref{Numerical experiments}) was produced, and plotted against the simulations fitted to the experimental data for illustration purposes.

Since all initial conditions have led to approximately the same ratio in each strain, the decision was made to produce the curves when \textsl{S. aureus} populations represented the resident populations in the numerical experiments in this study. This decision was based on two reasons. First, in this situation, SH1000 struggled the most to develop resistance. As previously indicated by the results of the interactions, negative correlations were observed between the initial density of the evolved \textsl{S. aureus} and its ability to recover and gain control over its opponents. Second, in this situation, when varying the ratio of the adapted SH1000, it is more evident when the resident populations can restrict and inhibit the invasions by \textsl{S. epidermidis} than under any other initial conditions. 

The models in this study predict that the final sample of evolved isolates of SH1000 will become resistant to the toxins produced by \textsl{S. epidermidis}, meaning that if samples were taken from the evolved SH1000 and engaged again in interactions with the same toxin-producing strains, the initial decline in SH1000 density would not be observed.

Thus, performing these numerical experiments aims to validate the model hypothesis of this study and to determine the proportion of evolved \textsl{S. aureus} SH1000 which must adapt to the toxins produced by \textsl{S. epidermidis} strains to be able to inhibit and restrict their invasions.

The numerical experiments were begun by assuming the reverse ratio between the fractions of \textsl{S. aureus} populations, i.e., the susceptible fraction represents 1\% of the evolved SH1000. Then the proportion of adaptive to susceptible was gradually reduced at a rate of ten percent until arriving at the initial assumption (light blue lines). However, only the following ($u_a: u_s$) ratios are shown in Fig (\ref{Numerical experiments}): 1\% light blue lines, 10\% orange lines, 30\% red lines, 50\% black lines, 60\% green lines, and 70\% dark red lines, for illustration purposes only. 

The outcomes of the numerical experiments, shown in Fig (\ref{testB180}), revealed that for the evolved \textsl{S. aureus} to restrict the invasions by B180 and recover, at least 50\% of \textsl{S. aureus} must mutate (black lines).

According to Fig (\ref{Three-variable model simulation Ch3 b}), mutation occurred at approximately day 4. Furthermore, when the ratio of the adaptive fraction of the \textsl{S. aureus} population was gradually reduced, the initial decline in the curves representing the density of \textsl{S. aureus} was observed, signifying that the evolved isolates of SH1000 were still sensitive to the toxins produced by B180, which contradicts the experimental findings. According to the results obtained from the deferred inhibition assay, in Fig (\ref{inhibition assay - after 2}), performed between the evolved populations of B180 and SH1000, B180 displayed no growth inhibition activity against the selected \textsl{S. aureus} SH1000. This indicates that by the end of interactions performed between SH1000 and B180 (day 17), the entire population of SH1000 changed and became resistant to the toxins produced by B180. 

Similarly, as illustrated in Fig (\ref{testB155}), the numerical experiments demonstrated that when the interactions were conducted between SH1000 and B155, at least 60\% or above of the evolved SH1000 had to adapt against the toxins produced by B155 before the evolution of the resistance became evident and recognised, which explains the delay in the process. According to Fig (\ref{Four-variable model simulation CH3 b}), this occurred between days 11 and 12. Also, as shown in Fig (\ref{Four-variable model simulation CH3 b}), by the end of interactions performed between SH1000 and B155 (day 28), the majority of the evolved SH1000 changed and became resistant to the toxins produced by B155. This was also confirmed by the experimental outcomes obtained in Fig (\ref{inhibition assay - after 2}).  

According to the results obtained from the deferred inhibition assay performed between the evolved populations of B155 and SH1000, B155 displayed no growth inhibition activity against the selected \textsl{S. aureus} SH1000 in some replicates. At the same time, it showed minor inhibition activity against the evolved SH1000 compared to the initial inhibition zones between the ancestral populations of B155 and SH1000 in other replicates. Likewise, as displayed in Fig (\ref{testB155}), when the ratio of the adaptive fraction of \textsl{S. aureus} was reduced to under 60\%, the initial decline in the curves representing the density of \textsl{S. aureus} was observed. This contradicts the experimental findings illustrated in Fig (\ref{C4}), which indicate that the change in the evolved population density is no longer significant regardless of the maintained oscillations.

Finally, as depicted in Fig (\ref{testTU3298}), the numerical experiments demonstrated that almost 70\% or above of the evolved SH1000 adapted against the toxins produced by TU3298 before the evolution of the resistance became evident (70\%-dark red, 80\%-yellow, 90\%-green, and 99\%-gray lines). According to Fig (\ref{Four-variable model simulation Ch3 b}), this happened approximately on days 9 and 10. According to the results obtained from the deferred inhibition assay performed between these evolved populations at day 28, some replicates showed no growth inhibition activity against the selected \textsl{S. aureus} SH1000. 

The numerical experiments demonstrated a positive correlation between the toxicity level of the reactions and the required resistance ratio. Furthermore, an inverse relationship was also noticed between the time required for the adaptive fraction of \textsl{S. aureus} to overcome the susceptible ratio and the growth rate of their opponent strain of \textsl{S. epidermidis}, i.e., the faster the evolved \textsl{S. epidermidis} strain grew, the more they provoked their opponent to transform. According to (Table \ref{tab:doubling time}), B180 had the highest growth rate out of the selected \textsl{S. epidermidis} strains, and as shown in Fig (\ref{Three-variable model simulation Ch33 b}), they were able to achieve the required ratio for the beginning of the resistance around the fourth day. In contrast, the most extended period consumed by the evolved SH1000 to reach the required ratio for the beginning of the resistance to evolve was when it participated in competitions with B155, which had the lowest growth rate.

Remarkably, while high toxin interactions urged their competitors from SH1000 to develop a higher resistance ratio, some of their evolved isolates showed minor inhibition activity against the evolved isolates of SH1000 at the end of interactions and maintained their existence at a higher ratio than the low-toxin strains. Such findings imply that the toxin-producer strains of \textsl{S. epidermidis} co-evolved in response to the survival challenges raised by increasingly resistant \textsl{S. aureus} populations. The evolved \textsl{S. epidermidis} may have increased the production of the inhibitory toxin or initiated the production of other toxins. However, this remains undetermined in the absence of an understanding of the inhibition mechanism.


{\textbf{\section{Conclusions}}}

In this study, a comprehensive study was presented that dealt with biological and mathematical aspects of the nature of the interactions in bacterial communities and the implications and consequences of these interactions. Several factors were considered when performing these experiments that could affect the dynamics of these interferences, for instance, the initial concentrations of the competing populations and the level of toxicity of the involved populations. Laboratory experiments were conducted where populations of the pathogenic strain \textsl{S. aureus}, SH1000, were engaged in competitions with different species of \textsl{S. epidermidis} that were distinguished by their level of toxicity.

The primary purpose of conducting these experiments was to investigate a hypothesis. Several studies confirm that the interactions between bacterial communities limit the colonisation of pathogenic bacteria \cite{A}. Moreover, studies have demonstrated that manipulating some of the environmental factors surrounding these interactions may contribute to the inhibition of pathogenic bacteria. Thus, several experiments were conducted to explore these hypotheses.

Three of the species used in this study were selected from a previous study presented in \cite{A}. Another species with a high toxicity level was added. A series of experiments (Table \ref{tab:experiments}) was conducted prior to the competition to gain a better understanding of the nature, characteristics, and features of these species, as well as to determine the extent of the change that may occur as a result of the competition. 

It was possible to accurately determine the production rate for each involved strain by incubating several replicates of each strain overnight and taking the $OD_{600}$ readings every 30 minutes. Inhibition assay experiments enabled the quantifying of the effect of bacteria-derived antimicrobials on competition. Incubating many replicates of the ($50 \mu l$) spot for each strain overnight allowed for the determination of the diffusion coefficients for all the involved strains. 

The principal component of this study was achieved by conducting competitions between the selected \textsl{S. epidermidis} species and the pathogenic \textsl{S. aureus} strain. Some factors were manipulated to ascertain their impact on the nature of these interactions, for example, the level of toxicity and the initial concentrations. These competitions were achieved from different initial concentrations, i.e., 0.01 : 1 and vice versa. These interactions were known as invasions, and they were also conducted from equal initial concentrations. These interactions lasted for different periods until the change of the competing population density was no longer significant. 

The most important results that were obtained through the study presented here is that the pathogenic species, SH1000, managed to dominate in all the competitions that were conducted in mixed environments, regardless of the toxicity of the competitor or the initial concentrations. This is consistent with the findings of the study presented in \cite{A}, in which it is stated that \textsl{Staphylococcus aureus} was only able to invade toxin-producing \textsl{S. epidermidis} under mixed conditions. 

The findings revealed the high adaptability of the pathogenic strain, SH1000, as it showed a decrease in its growth level at the beginning of these competitions. This decrease was positively proportional to the level of toxicity and the growth rate of the corresponding species. There is also a positive correlation between the level of toxicity of the competing strain of \textsl{S. epidermidis} and the time required for \textsl{S. aureus} to adapt and mutate against these toxins. The species with the highest toxicity was able to inhibit this species for a more extended period. 

Additionally, when conducting competitions with bacteria with a low level of toxicity, B180, toxins did not play a significant role in the outcomes of these interactions, when the interaction dynamics between SH1000 and B180 were compared with the interaction dynamics between the same species and another non-toxic \textsl{S. epidermidis} species, (from a previous study \cite{A}). 

Furthermore, a difference in the dynamics of these interferences was noticed when starting from different concentrations, as they showed clear and intense fluctuations during these competitions than when starting from equal concentrations (See Fig \ref{Fig:f10d}).  

None of the invasions and competitions performed and presented in this study resulted in any exclusions or total eliminations of the competitors, which differs from the findings in the study cited in \cite{A}. The evolved \textsl{S. epidermidis} coexisted with the opponent and was able to survive and maintain its presence at low concentrations.

To determine whether the emergence of genetic mutations is the reason behind the survival of this pathogenic species, a final experiment was conducted to measure the level of sensitivity of the evolved \textsl{S. aureus} populations when applied upon pure and evolved populations of \textsl{S. epidermidis}. This experiment showed that the pathogenic species were no longer affected by the toxins secreted by their opponents. In addition, it was noticed that the change was not limited to the pathogenic species but also affected the species that secrete toxins, as they became more ferocious when tested against pure samples of the pathogenic species (See figures \ref{inhibition assay - after 1} and \ref{inhibition assay - after 2}), signifying that both evolved species developed to suit the surrounding conditions and used all the tools that enable them to survive.

Mathematically, different models were designed to analyse and simulate each experiment performed. The first mathematical model was to simulate growth curves, and this model was based on the logistic equation where a single population consumes a single and limited resource. From this model, it was possible to estimate the consumption rate of each population along with their carrying capacity. In addition, the use of the logarithmic scale of the log phase in the obtained growth curves of the involved strains, (See Fig \ref{growth curves 2}), made it possible to accurately define doubling and relaxation times along with the growth rate for each population in minutes, hours, and days. By scaling the time in the MATLAB simulations to $t = 1$ is one day in our time unit, the findings helped to define the daily growth rate (Table \ref{tab:doubling time}) for the evolved species.

The outcomes of toxin-mediated inhibition experiments, which were performed to test the sensitivity of the pathogenic species to toxins secreted by other competing species, were modelled. Through this system of equations (\ref{toxin model}), it was possible to solve the toxin variable equation, (\ref{T_solution}), and obtain a toxin profile, Fig (\ref{T_profile}), as well as a mathematical term, (\ref{f_1 and f_2 equation}), that links the rate of secretion of toxins and the rate of their decrease, $f_1$ and $f_2$, with the inhibition coefficients, $p_1$ and $p_2$, that affect the susceptible and resistant fractions of \textsl{S. aureus} species. Such an approach was useful as there was no other way to estimate or determine the exact values for these parameters. 

The research presented in this study principally concerns modelling the interference between bacterial communities. These interactions were divided into two types. The first type was a competition for resources only, performed between B180 and SH1000, where toxicity did not play a significant role in the outcomes of these competitions; the other type was concerned with modelling the inhibitory interactions performed between samples of B155, TU3298 on the one hand, and SH1000 on the other. 

In the first type of these interactions, competition for resources, a three-variable model was developed to simulate the dynamics of interference. This model \eqref{system 2} was an expansion of the spatially homogeneous \textsl{Lotka-Volterra} competition model to include the diffusive terms of the evolved species. 
Observing the interactions between \textsl{S. aureus} and \textsl{S. epidermidis} a repeated pattern was detected where the pathogenic bacteria community begins to decline at the start of each competition and then persists to increase its population size. It was possible to replicate the non-monotonic behaviour seen in the interaction dynamics by assuming that the pathogenic bacteria, SH1000, comprises two fractions of the population. The susceptible fraction represents most of the population, and its inability to resist its opponents causes its decrease at the start of all interactions. 

In contrast, the resistant fraction represents only one per cent of the total population and can restructure the pathogen species, allowing it to increase production and dominate. Thus, a three-variable model was presented to consider the adapted fraction of \textsl{S. aureus} populations and its influence on the dynamics of interactions. A three-variable model was able to improve the simulations of the interference dynamics. However, this model could not produce the oscillations occurring in both populations during the competitions.

A mathematical model was also developed to simulate the inhibitory interaction dynamics in bacterial communities, considering the toxicity factor and its impact on the course of this dynamic. A four-variable model \eqref{system3} was introduced, where the toxin variable was added to the previous model and linked to the producers. The associated inhibition coefficients with the toxin variable on the \textsl{S. aureus} community with both parts were added. This model was more complex than the previous models due to its many parameters. However, the existence of a relationship between these parameters (\ref{f_1 and f_2 equation}) allowed the programming obstacle to be overcome in which the secretion factor, $f_1$, was set to one. Simultaneously, after defining the inhibition coefficients using their connection to $f_2$, the least square method was utilised to obtain the values with the least error for each interference coefficient and the rate of toxin dissolution. 

When modelling the interactions between \textsl{S. aureus} and \textsl{S. epidermidis} populations, the models predict the disappearance and demise of the B180 species after a period. At the same time, the competing communities that possess the toxic factors, B155 and TU3298, can survive at low rates if the resources are replenished. Moreover, the appropriate environment is maintained for their survival. 

In addition, according to the numerical experiments performed to test the modelling predictions, the final sample of evolved isolates of SH1000 become resistant to the toxins produced by \textsl{S. epidermidis}. There is a positive correlation between the toxicity level of the interactions and the required resistance ratio. Also, an inverse relationship was observed between the time required for the adaptive fracture of \textsl{S. aureus} to overcome the susceptible ratio and the growth rate of their opponent strain of \textsl{S. epidermidis}, i.e., the faster the evolved \textsl{S. epidermidis} strain grew, the more it provoked its opponent to develop the resistance. 

From the numerical experiments, also indicated in Fig (\ref{inhibition assay - after 2}), all the evolved populations of \textsl{S. epidermidis} changed to keep up with \textsl{S. aureus} species. This change was evident and observed in the evolved populations when they produced larger inhibition zones against the ancestral SH1000 compared to the inhibition zones created by the ancestral populations. 

Such findings may motivate and inspire future work, as it may be possible to expand the models in this study to consider the alterations in the evolved \textsl{S. epidermidis} populations, which are possibly the leading cause of the observed oscillations during the interference process. For example, another variable could be added to represent and express the change in these populations. Thus, both competing species can mutate and co-evolve to address the survival difficulties presented by their opponents.

{\textbf{\section*{Appendix}}}
\begin{figure}[H]
\centering
\begin{subfigure}{0.49\linewidth}
\centering
\caption{}
\label{B180_1-1}
\includegraphics[width=0.9\textwidth]{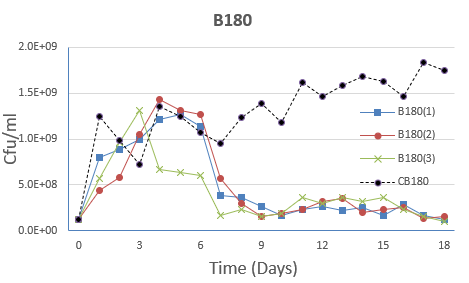}
\end{subfigure}
\hfill
\begin{subfigure}{0.49\linewidth}
\centering
\caption{}
\label{SH1000_B180_1-1}
\includegraphics[width=0.9\textwidth]{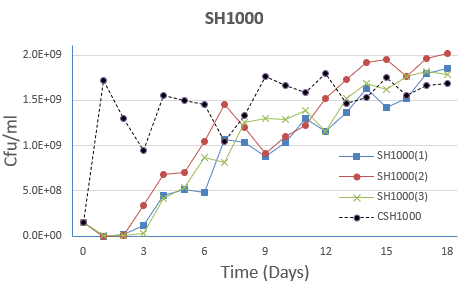}
\end{subfigure}
\caption{\textbf{Isolates of \textsl{S. aureus} (SH$1000$) competing with populations of low-toxin-producing \textsl{S. epidermidis} (B$180$) at initial frequencies of $1:1$.}\, Panel (a): shows the initial data obtained from three replicates, representing the behaviour of the evolved populations B$180$.  Panel (b): the behaviour of the evolved populations SH$1000$, along with the controls (dotted line), where the populations were cultured independently. The $x$-axis is the time in days, and the $y$-axis is the colony-forming units (cfu) per plate.}
\label{control control}
\end{figure}
\begin{figure}[H]
\centering
\begin{subfigure}{0.49\linewidth}
\centering
\caption{}
\label{}
\includegraphics[width=0.7\textwidth]{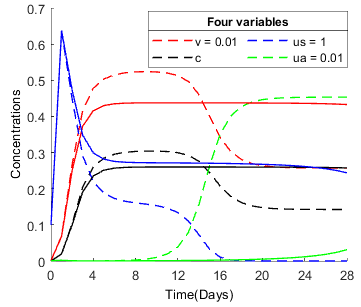}
\end{subfigure}
\hfill
\begin{subfigure}{0.49\linewidth}
\centering
\caption{}
\label{}
\includegraphics[width=0.7\textwidth]{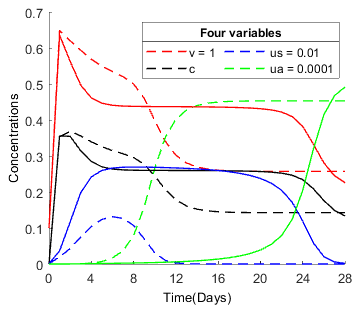}
\end{subfigure}
\caption{\textbf{The effect of different inhibition coefficients on the dynamics of interactions.}\,(a). Simulations of interactions when a toxin-producing population of \textsl{Staphylococcus epidermidis} TU$3298$ invades a resident population of \textsl{Staphylococcus aureus} SH$1000$ at a concentration of $0.01:1$. (b). Simulations of interactions in the opposite direction. Solid lines indicate when the inhibition coefficients are lower than the dashed lines. Common parameters: $D_u=1.8\times\,10^{-6}$, $D_v=5\times\,10^{-5}$, $r_u=26.5296$, $r_v=28.613$, $b_1=0.9$, $b_2=0.82$, $b_3=0.9$ and $b_4=0.999$. Solid lines parameters: $p_1=0.0968$, $p_2=0.0966$, $f_1=1$, $f_2=0.03$. Dashed lines parameters: $p_1=0.2258$, $p_2=0.2247$, $f_1=1$, $f_2=0.07$.}
\label{different_fs}
\end{figure}

\end{document}